\documentclass[twocolumn]{aastex63}

\usepackage{color}
\usepackage{natbib}
\usepackage{graphicx}
\usepackage{float}
\usepackage{amsmath}
\usepackage{amssymb}
\usepackage{bm}
\usepackage[toc,page]{appendix}
\usepackage{multirow}

\def\ra#1#2#3{#1$^{\rm h}$ #2$^{\rm m}$ #3$^{\rm s}$}
\def\dec#1#2#3{$#1^\circ #2' #3''$}

\shortauthors{Schroeder et al.}
\shorttitle{The afterglow of short GRB 210726A}

\begin{document}

\author[0000-0001-9915-8147]{Genevieve~Schroeder}
\affiliation{Center for Interdisciplinary Exploration and Research in Astrophysics (CIERA) and Department of Physics and Astronomy, Northwestern University, Evanston, IL 60208, USA}

\author[0000-0003-2705-4941]{Lauren~Rhodes}
\affiliation{Astrophysics, Department of Physics, University of Oxford, Keble Road, Oxford, OX1 3RH, UK}

\author[0000-0003-1792-2338]{Tanmoy Laskar}
\affiliation{Department of Physics \& Astronomy, University of Utah, Salt Lake City, UT 84112, USA}
\affiliation{Department of Astrophysics/IMAPP, Radboud University, P.O. Box 9010, 6500 GL, Nijmegen, The Netherlands}

\author[0000-0002-2028-9329]{Anya Nugent}
\affiliation{Center for Interdisciplinary Exploration and Research in Astrophysics (CIERA) and Department of Physics and Astronomy, Northwestern University, Evanston, IL 60208, USA}

\author[0000-0003-3937-0618]{Alicia Rouco Escorial}
\affiliation{European Space Agency (ESA), European Space Astronomy Centre (ESAC), Camino Bajo del Castillo s/n, 28692 Villanueva de la Cañada, Madrid, Spain}

\author[0000-0002-9267-6213]{Jillian C.~Rastinejad}
\affiliation{Center for Interdisciplinary Exploration and Research in Astrophysics (CIERA) and Department of Physics and Astronomy, Northwestern University, Evanston, IL 60208, USA}

\author[0000-0002-7374-935X]{Wen-fai Fong}
\affiliation{Center for Interdisciplinary Exploration and Research in Astrophysics (CIERA) and Department of Physics and Astronomy, Northwestern University, Evanston, IL 60208, USA}

\author[0000-0001-9149-6707]{Alexander J. van der Horst}
\affiliation{Department of Physics, The George Washington University, 725 21\textsuperscript{st} Street NW, Washington DC, 20052, USA}

\author[0000-0002-2149-9846]{P\'eter~Veres}
\affiliation{Department of Space Science, University of Alabama in Huntsville, Huntsville, AL 35899, USA}
\affiliation{Center for Space Plasma and Aeronomic Research, University of Alabama in Huntsville, Huntsville, AL 35899, USA}

\author[0000-0002-8297-2473]{Kate D.~Alexander}
\affiliation{Steward Observatory, University of Arizona, 933 North Cherry Avenue, Tucson, AZ 85721-0065, USA}

\author[0000-0003-2734-1895]{Alex Andersson}
\affiliation{Astrophysics, Department of Physics, University of Oxford, Keble Road, Oxford, OX1 3RH, UK}

\author[0000-0002-9392-9681]{Edo Berger}
\affiliation{Center for Astrophysics | Harvard \& Smithsonian, Cambridge, MA 02138, USA}

\author[0000-0003-0526-2248]{Peter K. Blanchard}
\affiliation{Center for Interdisciplinary Exploration and Research in Astrophysics (CIERA) and Department of Physics and Astronomy, Northwestern University, Evanston, IL 60208, USA}

\author[0000-0003-3507-335X]{Sarah Chastain}
\affiliation{Department of Physics and Astronomy, University of New Mexico, 210 Yale Blvd NE, Albuquerque, NM 87106, USA}

\author[0000-0001-8415-7547]{Lise Christensen}
\affiliation{Cosmic Dawn Center (DAWN), University of Copenhagen, Denmark}
\affiliation{Niels Bohr Institute, University of Copenhagen, Jagtvej 128, DK-2200 N, Copenhagen, Denmark}

\author{Rob Fender}
\affiliation{Astrophysics, Department of Physics, University of Oxford, Keble Road, Oxford, OX1 3RH, UK}
\affiliation{Department of Astrophysics, University of Cape Town, Private Bag X3, Rondebosch, Cape Town, 7701, South Africa}

\author[0000-0003-3189-9998]{David A.\ Green}
\affiliation{Cavendish Laboratory, University of Cambridge, 19 J.~J .\ Thomson Avenue, Cambridge CB3 0HE, UK}

\author[0000-0002-4488-726X]{Paul Groot}
\affiliation{Department of Astrophysics, University of Cape Town, Private Bag X3, Rondebosch, Cape Town, 7701, South Africa}
\affiliation{South African Astronomical Observatory, PO Box 9, Observatory 7935, South Africa}
\affiliation{Department of Astrophysics/IMAPP, Radboud University, P.O. Box 9010, 6500 GL, Nijmegen, The Netherlands}

\author[0000-0001-6864-5057]{Ian Heywood}
\affiliation{Astrophysics, Department of Physics, University of Oxford, Keble Road, Oxford, OX1 3RH, UK}
\affiliation{Department of Physics and Electronics, Rhodes University, PO Box 94, Makhanda 6140, South Africa}
\affiliation{South African Radio Astronomy Observatory (SARAO), 2 Fir Street, Observatory, Cape Town 7925, South Africa}

\author[0000-0002-5936-1156]{Assaf Horesh}
\affiliation{Racah Institute of Physics, The Hebrew University of Jerusalem, Jerusalem, 91904, Israel}

\author[0000-0001-9695-8472]{Luca Izzo}
\affiliation{DARK, Niels Bohr Institute, University of Copenhagen, Jagtvej 128, DK2200 Copenhagen, Denmark}

\author[0000-0002-5740-7747]{Charles~D.~Kilpatrick}
\affiliation{Center for Interdisciplinary Exploration and Research in Astrophysics (CIERA) and Department of Physics and Astronomy, Northwestern University, Evanston, IL 60208, USA}

\author{Elmar K\"{o}rding}
\affiliation{Department of Astrophysics/IMAPP, Radboud University, P.O. Box 9010, 6500 GL, Nijmegen, The Netherlands}

\author[0000-0002-7851-9756]{Amy Lien}\affiliation{University of Tampa, Department of Physics and Astronomy, 401 W. Kennedy Blvd, Tampa, FL 33606, USA}

\author[0000-0002-7517-326X]{Daniele B. Malesani}
\affiliation{Department of Astrophysics/IMAPP, Radboud University, P.O. Box 9010, 6500 GL, Nijmegen, The Netherlands}\affiliation{Cosmic Dawn Center (DAWN), Denmark}\affiliation{Niels Bohr Institute, University of Copenhagen, Jagtvej 128, 2200 Copenhagen N, Denmark.}

\author[0000-0002-8133-3100]{Vanessa McBride}
\affiliation{Department of Astrophysics, University of Cape Town, Private Bag X3, Rondebosch, Cape Town, 7701, South Africa}
\affiliation{South African Astronomical Observatory, PO Box 9, Observatory 7935, South Africa}
\affiliation{IAU Office of Astronomy for Development, P.O. Box 9, 7935 Observatory, South Africa}

\author[0000-0002-2557-5180]{Kunal Mooley}
\affiliation{National Radio Astronomy Observatory, Socorro, 87801, New Mexico, USA}
\affiliation{Cahill Center for Astronomy and Astrophysics, California Institute of Technology, Pasadena, 91125, CA, USA}

\author[0000-0002-1195-7022]{Antonia Rowlinson}
\affiliation{Anton Pannekoek Institute for Astronomy, University of Amsterdam, Science Park 904, NL-1098 XH Amsterdam, The Netherlands}
\affiliation{ASTRON, the Netherlands Institute for Radio Astronomy, Postbus 2, NL-7990 AA Dwingeloo, The Netherlands}

\author[0000-0001-8023-4912]{Huei Sears}
\affiliation{Center for Interdisciplinary Exploration and Research in Astrophysics (CIERA) and Department of Physics and Astronomy, Northwestern University, Evanston, IL 60208, USA}

\author[0000-0001-9242-7041]{Ben Stappers}
\affiliation{Jodrell Bank Centre for Astrophysics, Department of Physics and Astronomy, The University of Manchester, Manchester, M13 9PL, UK}

\author[0000-0003-3274-6336]{Nial Tanvir}
\affiliation{School of Physics and Astronomy, University of Leicester, University Road, Leicester, LE1 7RH, UK}

\author[0000-0001-9398-4907]{Susanna D. Vergani}
\affiliation{GEPI, Observatoire de Paris, Université PSL, CNRS, 5 Place Jules Janssen, 92190 Meudon, France}

\author[0000-0002-3101-1808]{Ralph A.M.J. Wijers}
\affiliation{Anton Pannekoek Institute for Astronomy, University of Amsterdam, Science Park 904, NL-1098 XH Amsterdam, The Netherlands}

\author[0000-0001-7361-0246]{David Williams-Baldwin}
\affiliation{Jodrell Bank Centre for Astrophysics, School of Physics and Astronomy, The University of Manchester, Manchester, M13 9PL, UK}

\author[0000-0002-6896-1655]{Patrick Woudt}
\affiliation{Department of Astrophysics, University of Cape Town, Private Bag X3, Rondebosch, Cape Town, 7701, South Africa}

\title{A Radio Flare in the Long-Lived Afterglow of the Distant Short GRB 210726A: Energy Injection or a Reverse Shock from Shell Collisions?}

\begin{abstract}
    We present the discovery of the radio afterglow of the short $\gamma$-ray burst (GRB) 210726A, localized to a galaxy at a photometric redshift of $z\sim 2.4$. While radio observations commenced $\lesssim 1~$day after the burst, no radio emission was detected until $\sim11$~days. The radio afterglow subsequently brightened by a factor of $\sim 3$ in the span of a week, followed by a rapid decay (a ``radio flare''). We find that a forward shock afterglow model cannot self-consistently describe the multi-wavelength X-ray and radio data, and underpredicts the flux of the radio flare by a factor of $\approx 5$. We find that the addition of substantial energy injection, which increases the isotropic kinetic energy of the burst by a factor of $\approx 4$, or a reverse shock from a shell collision are viable solutions to match the broad-band behavior. At $z\sim 2.4$, GRB\,210726A is among the highest redshift short GRBs discovered to date as well as the most luminous in radio and X-rays. Combining and comparing all previous radio afterglow observations of short GRBs, we find that the majority of published radio searches conclude by $\lesssim 10~$days after the burst, potentially missing these late rising, luminous radio afterglows.
\end{abstract}

\section{Introduction}

Short duration $\gamma$-ray bursts (GRBs) are cosmological explosions produced in compact object mergers involving neutron stars \citep[NS, ][]{Eichler1989Natur.340..126E_elps89, Narayan1992ApJ...395L..83N_npp1992, Berger2014ARA&A..52...43B, abbott2017}. 
The interaction of the GRB jet with the circumburst environment produces a synchrotron ``afterglow", resulting in detectable emission from the X-ray to the radio bands \citep[i.e.,][]{spn1998_sari, Wijers1999ApJ...523..177WG1999, GS2002}. The afterglow peak moves from the optical to radio wavebands as the jet expands and evolves, indicating that radio frequency observations provide the best opportunity to detect the afterglow for the longest time. The behavior of the radio afterglow is also a strong function of the circumburst environment and jet microphysics. 

Radio detections of short GRB afterglows, in combination with X-ray and optical observations, can reveal or constrain behavior that deviates from the standard forward shock afterglow model and the true energetics of the burst. For instance, radio detections of short GRB afterglows have helped reveal the the presence of energy injection and reverse shocks (RSs), as well as constrain the jet opening angles and extreme energetics  \citep{sbk+2006ApJ...650..261S, fbm+2014ApJ...780..118F, tsc+2016ApJ...827..102T, ltl+2019ApJ...883...48L, tctb+2019MNRAS.489.2104T, flr+2021ApJ...906..127F, lers+2022}. Importantly, without the radio detections, these short GRBs may have been considered ``typical" within the population based on their prompt or X-ray/optical properties alone.

While the majority of short GRBs discovered by NASA's Neil Gehrels {\it Swift} Observatory ({\it Swift}) have detected X-ray afterglows \citep[$\approx 70\%$, ][]{fbm+2015, refb+2022}, only thirteen ($\approx 11\%$) have detected radio afterglows \citep{BPC+2005Natur.438..988B, sbk+2006ApJ...650..261S, fbm+2014ApJ...780..118F, fbm+2015, 2017GCN.21395....1F, ltl+2019ApJ...883...48L, flr+2021ApJ...906..127F, 2021GCN.30658....1S_VLAdet, lers+2022, 2023GCN.33372....1S_230205A, 2023GCN.33358....1S_230217A}. This is in part due to their low beaming-corrected kinetic energies compared to their long-duration counterparts($\approx 10^{51}~$erg) and generally low circumburst densities \citep[$\approx 10^{-1}~{\rm cm}^{-3}$, ][]{2006MNRAS.367L..42P, sbk+2006ApJ...650..261S, Gehrels2008, fbm+2015,  refb+2022}. In addition, radio observations of short GRBs have primarily concluded at $\lesssim 10~$days post-burst \citep{flr+2021ApJ...906..127F}, possibly missing the peak in the light curve of the radio afterglow ($\sim 6$--$20$)~days for typical explosion properties. 
Indeed, the radio emission could still be rising at $\gtrsim 10~$days post-burst, especially for GRBs with a wide and energetic jet, such as GRB\,211106A \citep{lers+2022}. Therefore, it is clear that radio observations extending to later times have the potential to reveal short GRBs with wider jets, larger energy scales, and/or additional emission mechanisms.

Here, we present our multi-wavelength monitoring of the afterglow of the short-duration GRB\,210726A, including a rapidly rising and fading radio afterglow first detected at $\sim 11~$days post-burst. In Section~\ref{sec:observations} we present the burst discovery, and X-ray, optical and radio observations. In Section~\ref{sec:HostGalaxy}, we present host galaxy observations, spectral energy distribution (SED) modeling and the derived photometric redshift. In Sections~\ref{sec:afterglow}-\ref{sec:fs_model} we explore and fit the temporal and spectral evolution of the afterglow to a standard forward shock (FS) synchrotron model to determine burst explosion properties. We further establish the existence of a radio flare in the 3-10~GHz light curves which cannot be explained by the FS model alone. We explore alternative scenarios to the standard FS model in Section~\ref{sec:Alternative_theories}, with energy injection or a reverse shock (RS) emerging as the possible causes of the radio flare. In Section~\ref{sec:Discussion} we discuss the implications of energy injection and RSs in short GRBs, provide a comparison to other short GRBs, and reflect on future radio monitoring of short GRB afterglows. We conclude in Section~\ref{sec:Conclusions}. In this paper, we employ the $\Lambda$CDM cosmological parameters of $H_{0} = 68~{\rm km \, s}^{-1} \, {\rm Mpc}^{-1}$, $\Omega_{M} = 0.31$, $\Omega_{\rm \Lambda} = 0.69$.

\section{Burst and Afterglow Observations}
\label{sec:observations}

\subsection[gamma-ray Detection]{$\gamma$-ray Detection}
\subsubsection{Swift}

GRB\,210726A was discovered by the Burst Alert Telescope (BAT; \citealt{Swift_BAT_2005SSRv..120..143B}) on-board the Neil Gehrels {\it Swift} Observatory ({\it Swift}; \citealt{Swift_2004ApJ...611.1005G}) on 2021 July 26 at 19:19:03 UT \citep{2021GCN.30523....1B_Swiftdet}. The burst duration was determined to be $T_{90} \approx 0.39~{\rm s}$ with a fluence of $f_{\gamma} ({\rm 15-150~keV}) = 4.3 \pm 1.1 \times 10^{-8}~{\rm erg~cm}^{-2}$ \citep{shortburst_2021GCN.30535....1T, 2021GCN.30536....1P_SwiftBATRefined}. The position of the burst was refined to R.A.=\ra{12}{53}{19.9} and decl.=\dec{+19}{12}{38.9} (J2000) with a positional uncertainty of $2.6'$ radius (90\% containment; \citealt{2021GCN.30536....1P_SwiftBATRefined}). We calculate a hardness ratio (${\rm HR}$) of ${\rm HR}_{\rm BAT} = f_{\gamma} ({\rm 50-100~keV})/f_{\gamma} ({\rm 25-50~keV}) \approx 1.4$ \citep{Lien2016}. When compared to other short GRBs in the hardness versus duration plane, GRB\,210726A is similar to other BAT-detected short GRBs\footnote{https://swift.gsfc.nasa.gov/results/BATbursts/1061687/bascript/\\top.html\#Dist}.

\subsubsection{Fermi}
\label{sec:Fermi_GBM}

While GRB\,210726A did not trigger the Gamma-ray Burst Monitor (GBM; \citealt{GBM_2009ApJ...702..791M}) on-board Fermi \citep{Fermi_1999APh....11..277G}, we used a {\tt targeted search} pipeline \citep{targetedsearch2019arXiv190312597G} to identify a transient source in the Fermi-GBM data, coincident in time and location with the {\it Swift}-BAT alert \citep{Fermi_det_2021GCN.30540....1V}. The source was detected most significantly on the 0.512 s timescale with a log-likelihood ratio (LLR) of $\sim 12.6$. This corresponds to a signal-to-noise ratio of $\sim 6$ \citep{2018ApJ...862..152K}. We note that using the {\tt targeted search}, known outside triggers (e.g. {\it Swift}-only triggers like this case) with LLR as low as 10 have been recovered in GBM data.

We find the best-fit model for the spectrum of GRB\,210726A ($\delta t = 0.736$--$1.376~$s, where $\delta t$ is the time after BAT trigger) is a simple power law with photon index $\Gamma=-1.88 \pm 0.21 $, and fluence $f_{\gamma}(10-1000 {\rm keV}) =(1.45\pm 0.47)\times10^{-7}$ erg cm$^{-2}$. A  power-law model with exponential cutoff fits the spectrum equally well, but it is not statistically preferred. We constrain the peak energy to $E_{\rm peak}=53\pm 18$ keV,  suggesting a GRB that is soft compared to other short GRBs.

We calculate an ${\rm HR}$ of ${\rm HR}_{\rm GBM}={\rm Counts (50-300~keV)}/{\rm Counts (10-50~keV)} = 0.28\pm0.18$ \citep{FERMI_2016ApJS..223...28N}, placing GRB\,210726A among the softest short GRBs detected by Fermi/GBM. This is in stark contrast to the calculated  ${\rm HR}_{\rm BAT} \sim 1.4$. The discrepancy lies in both the method and the energy ranges involved in calculating ${\rm HR}$ for BAT and GBM, where ${\rm HR}_{\rm BAT}$ is calculated using the fluence, rather than the counts, in narrower energy bands than GBM. Indeed, integrating the GBM power law spectrum and deriving the ${\rm HR}$ similarly to BAT (i.e. calculating $f_{\gamma} ({\rm 50-100~keV})/f_{\gamma} ({\rm 25-50~keV})$), we find ${\rm HR}_{\rm GBM \rightarrow BAT}=1.1\pm 0.3$, which is consistent with the measured value of ${\rm HR}_{\rm BAT} \approx 1.4$.

\subsection{X-rays}

\subsubsection{Swift}
\label{sec:Swift_Obs}

The {\it Swift}/X-ray Telescope (XRT) started observations of GRB\,210726A at $\delta t = 60.0$~s, finding an uncatalogued X-ray source within the BAT position \citep{2021GCN.30523....1B_Swiftdet, 2021GCN.30537....1D_SwiftXRTRefined}. The refined XRT position of the GRB is R.A.=\ra{12}{53}{09.82} decl.=\dec{+19}{11}{25.1} (J2000) with a positional uncertainty radius of $1.7\arcsec$ (90\% containment; \citealt{2021GCN.30524....1O_enhenceXRT,Evans2009}). The X-ray afterglow of GRB\,210726A was observed and detected by {\it Swift} until $\delta t \approx 5.1 \times 10^{5}~$s. We obtain all the available information of the X-ray burst afterglow from the {\it Swift} light curve repository\footnote{\url{https://www.swift.ac.uk/xrt_curves/}}.

\begin{figure*}
    \centering
    \includegraphics[width = 1.1\columnwidth]{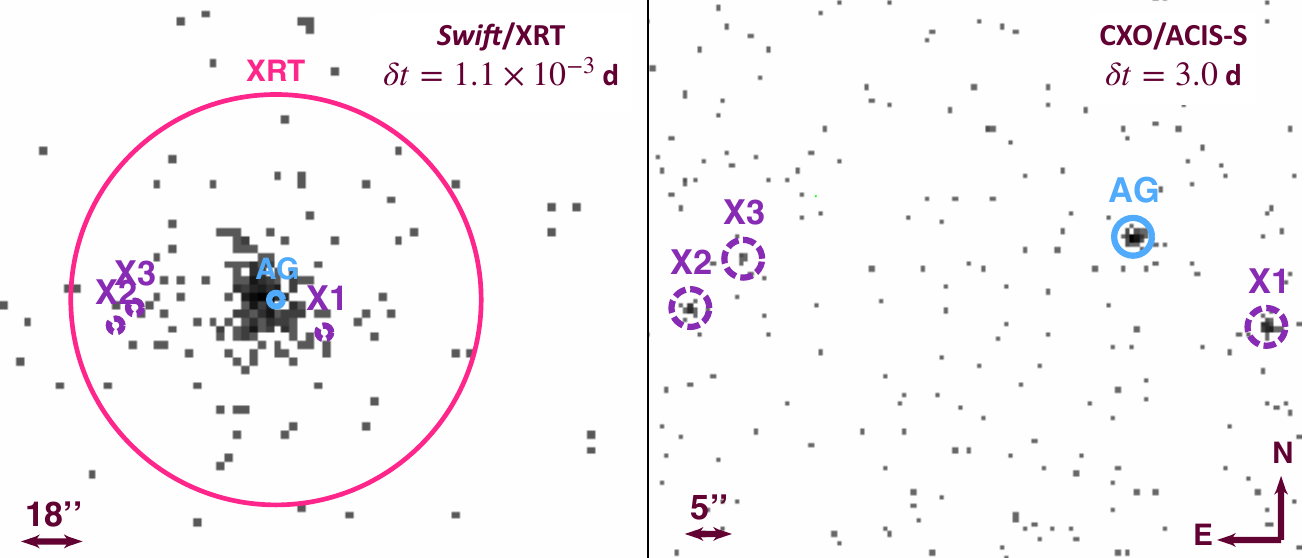}
    \includegraphics[width = 0.8\columnwidth]{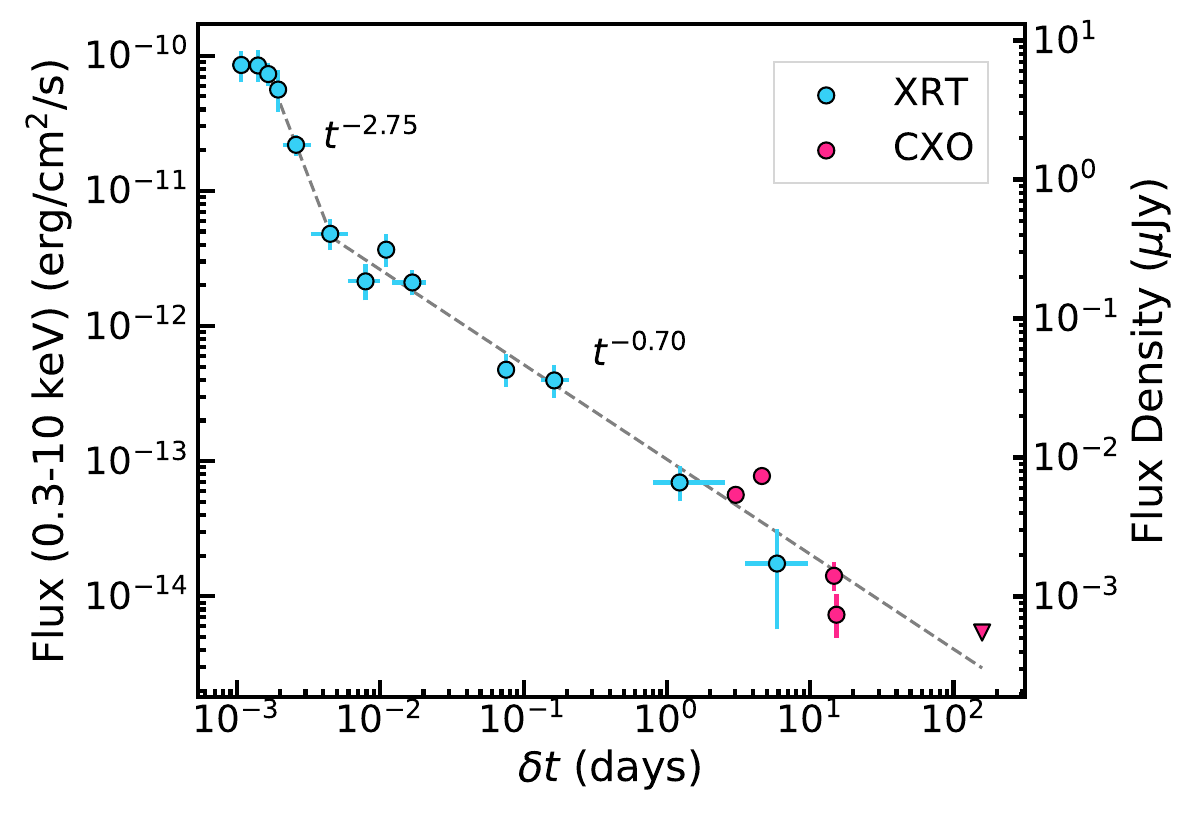}
    \caption{\emph{Left:} {\it Swift}/XRT and CXO/ACIS-S images of the field of GRB\,210726A in the $0.3-10$\,keV (left) and $0.5-8$\,keV (middle) energy bands, respectively. The large pink circle shows the XRT afterglow extraction region ($\sim 70''$) at early times, while the blue circle indicates the CXO source region. Contaminants contributing to the XRT flux (X1, X2 and X3) are labeled and shown as purple dashed regions. 
    \emph{Right:} The {\em Swift}/XRT and CXO afterglow light curve of GRB\,210726A. Dashed lines represent power law fits to the X-ray afterglow at ``early" time ($\delta t <(1.5$--$3.5) \times 10^{-3}~$days) and ``late" time ($\delta t > 3.5 \times 10^{-3}~$days) X-ray afterglows, which have temporal declines of $-2.75 \pm 0.09$ and $-0.70 \pm 0.05$, respectively (the latter is $-0.75 \pm 0.03$ if the re-brightening at $\approx4.6$~days is excluded).}
    \label{fig:xray_image_lc}
\end{figure*}

\subsubsection{Chandra}
\label{sec:Chandra_Obs}

We triggered the {\it Chandra} X-ray Observatory (CXO; \citealt{Chandra_2000SPIE.4012....2W}) (Program 22400461, ObsIDs: 23445, 23446, 25680, PI: Fong; Program 22500107, ObsID: 23542, PI: Berger) starting at $\delta t \approx 3.0~$days, and obtained a late-time follow-up observation (through DDT, Program 22408825, ObsID: 26247, PI: Schroeder) at $\delta t \approx 158.2~$days. For the CXO data reduction and subsequent analysis, we used the \texttt{CIAO} software package (v.\,4.12) with calibration files (\texttt{caldb}; v.\,4.9.0). We obtained new Level II event files after reprocessing the data utilizing \texttt{chandra\_repro}, and filtered the dataset against any high background activity. We detected the X-ray afterglow of GRB\,210276A \citep{chandra_2021GCN.30558....1R} using the ACIS (Advanced CCD Imaging Spectrometer) detector \citep{Garmire2003}, and continued to detect the burst afterglow until $\delta t \approx 15.2~$days, when the burst went into Sun constraint. We did not detect any X-ray emission at $\delta t \approx 158.2~$days (Figure\,\ref{fig:xray_image_lc}). 

We run the \texttt{CIAO} correcting absolute astrometry task\footnote{\url{https://cxc.cfa.harvard.edu/ciao/threads/reproject\_aspect/}} using the USBO-B1.0 optical catalog, and performed a blind search for X-ray sources with \texttt{CIAO}/\texttt{wavdetect} on the first CXO observation ($\sim 25$\,ks effective exposure time) at $\delta t \approx 3.0~$days. We determine a position of R.A.=\ra{12}{53}{09.81}, decl.=\dec{+19}{11}{25.3} (J2000) with a total positional uncertainty of $0.6\arcsec$ ($1 \sigma$).

During the blind search for X-ray sources, we also detected 3 neighboring sources (hereafter X1, X2 and X3) in all the CXO observations at $\sim 17''$, $\sim 62''$ and $\sim56''$, respectively, offset from the afterglow (X1-X3; Figure\,\ref{fig:xray_image_lc}). These sources are not distinguishable from the burst afterglow in the XRT observations due to the initial $\sim 70 \arcsec$-radius extraction region of XRT (\citealt{Evans2007, Evans2009}, Figure~\ref{fig:xray_image_lc}). Thus, to quantify and correct for the contaminant contributions to the XRT afterglow flux, we extracted the spectral information from the position of the afterglow and each contaminant across all CXO observations and accounted for them in the spectral analysis. As the afterglow fades, the XRT source region radius decreases from $\sim 70\arcsec$ to $\sim 21\arcsec$ \citep{Evans2007}, so we take into account which contaminant is present in the spectral analysis of each XRT bin. The details of this analysis are presented in Appendix~\ref{sec:Xray_contaminants}.

\subsubsection{Spectral analysis}
\label{sec:Xray_lc_spectra}

The XRT data spans $\delta t \approx 9.8\times 10^{-4}$~days to $\delta t \approx 9.2$~days, and the XRT light curve is dynamically\footnote{See: \url{https://www.swift.ac.uk/xrt_curves/docs.php\#products}} binned into 13 time bins. We generate the spectra for each bin ($0.3-10$\,keV) using the ``create time-sliced spectra'' option in the {\it Swift} repository.

\begin{deluxetable*}{ccccc}[t!]
 \centering
 \tabletypesize{\footnotesize}
 \tablecolumns{5}
 \tablecaption{X-ray observations of GRB\,210726A}
 \tablehead{   
   \colhead{Time} &
      \colhead{Bin length}&
   \colhead{Net counts}&
   \colhead{Unabsorbed Flux} &
   \colhead{Flux Density}\\[-0.1in]
   \colhead{(days)} &
   \colhead{(s)} &
   \colhead{} &
   \colhead{($10^{-14}$\,erg\,s$^{-1}$\,cm$^{-2}$)}&
   \colhead{($\mu$Jy)}
   }
\startdata 
\multicolumn{4}{c}{{\it Swift}/XRT} \\
\hline
$1.1 \times 10 ^{-3}$ & $3.0\times 10 ^{1}$  & 18  &  $(8.5_{-2.2}^{+2.3}) \times 10^{3}$  & $6.5 \pm 1.7$ \\
$1.4 \times 10 ^{-3}$ & $2.0\times 10 ^{1}$  & 15  &  $(8.5_{-2.1}^{+2.5}) \times 10^{3}$  & $6.5 \pm 1.7$ \\
$1.7 \times 10 ^{-3}$ & $2.5\times 10 ^{1}$  & 27  &  $(7.3_{-1.4}^{+1.6}) \times 10^{3}$  & $5.6 \pm 1.1$\\
$1.9 \times 10 ^{-3}$ & $2.5\times 10 ^{1}$  &  9  &  $(5.6_{-1.8}^{+2.2}) \times 10^{3}$  & $4.3 \pm 1.5$\\
$2.6 \times 10 ^{-3}$ & $1.0\times 10 ^{2}$  & 34  &  $(2.2\pm0.4) \times 10^{3}$          & $1.7 \pm 0.3$\\
$4.5 \times 10 ^{-3}$ & $2.3\times 10 ^{2}$  & 15  &  $(4.8_{-1.2}^{+1.4}) \times 10^{2}$  & $(5.1 \pm 1.3)\times 10^{-1}$\\
$7.9 \times 10 ^{-3}$ & $3.5\times 10 ^{2}$  & 12  &  $(2.1_{-0.6}^{+0.7}) \times 10^{2}$  & $(2.2  \pm 0.7)\times 10^{-1}$\\
$1.1 \times 10 ^{-2}$ & $1.8\times 10 ^{2}$  & 15  & $(3.7_{-0.9}^{+1.1}) \times 10^{2}$   & $(3.9 \pm 1.1)\times 10^{-1}$\\
$1.7 \times 10 ^{-2}$ & $7.4\times 10 ^{2}$  & 26  &  $(2.1_{-0.4}^{+0.5}) \times 10^{2}$  & $(2.2 \pm 0.5)\times 10^{-1}$\\
$7.5 \times 10 ^{-2}$ & $1.7\times 10 ^{3}$  & 14  &  $(4.8_{-1.2}^{+1.4}) \times 10^{1}$  & $(5.0 \pm 1.4)\times 10^{-2}$\\
$1.6 \times 10 ^{-1}$ & $1.9\times 10 ^{3}$  & 16  &  $(4.0_{-1.0}^{+1.2}) \times 10^{1}$  & $(4.2 \pm 1.2)\times 10^{-2}$\\
1.2                             & $1.4\times 10 ^{4}$ & 28 & $7.0_{-1.9}^{+2.2}$         &   $(7.3 \pm 2.2)\times 10^{-3}$   \\
5.9$^{*}$                       & $2.3\times 10 ^{4}$ & 10 & $1.8_{-1.2}^{+1.4}$         &   $(1.8 \pm 1.4)\times 10^{-3}$   \\
\hline
\multicolumn{5}{c}{CXO/ACIS-S} \\
\hline
$3.0$   & $2.5\times 10 ^{4}$  & 76  & $5.6\pm0.7$   & $(5.9 \pm 0.7)\times 10^{-3}$\\
$4.6$   & $2.0\times 10 ^{4}$  & 78  &  $7.8_{-0.9}^{+1.0}$  & $(8.2 \pm 1.0)\times 10^{-3}$\\
$14.7$  & $2.5\times 10 ^{4}$  & 17  &  $1.4_{-0.3}^{+0.4}$  & $(1.5 \pm 0.4)\times 10^{-3}$\\
$15.2$  & $2.5\times 10 ^{4}$  & 8   &  $0.7_{-0.2}^{+0.3}$  & $(7.7 \pm 2.8)\times 10^{-4}$\\
$158.2$ & $3.5\times 10 ^{4}$  & 1   &  $<0.5$   & $< 5.7 \times 10^{-4}$ \\
\hline
\hline
\multicolumn{5}{c}{Best-fit Spectral Parameters} \\
\hline
\multicolumn{3}{c}{$\delta t$}     & \multicolumn{2}{c}{$\Gamma_{X}$}\\
\multicolumn{3}{c}{(days)}     &   \multicolumn{2}{c}{}              \\
\hline
\multicolumn{3}{c}{$\lesssim 3.5 \times 10^{-3}$ (early)}    &   \multicolumn{2}{c}{$1.5\pm0.2$}  \\
\multicolumn{3}{c}{$\gtrsim 3.5 \times 10^{-3}$ (late)}      &   \multicolumn{2}{c}{$1.8\pm0.1$}  \\
\enddata
\tablecomments{Time is log-centered (observer frame). Fluxes are reported in the 0.3--10\,keV band. Uncertainties are $1\sigma$\\
$^{*}$ We linked the spectral parameters of the last two bins to account for more counts and to improve the statistics for the spectral analysis.}
\label{tab:data_xray}
\end{deluxetable*}

To determine if there is spectral evolution of the afterglow, we jointly fit the {\it Swift}/XRT and CXO observations using a four-component model to account for the combined presence of the afterglow, X1, X2 and X3, combining one double and three single-absorbed power-law models as follows: \texttt{[tbabs x ztbabs x (const x pow)]$_{\text{AG}}$ + [tbabs x (const x pow)]$_{\text{X1}}$ + [tbabs x (const x pow)]$_{\text{X2}}$ + [tbabs x (const x pow)]$_{\text{X3}}$}. We use a \texttt{constant} multiplicative model to include the cross-calibration between {\it Swift}/XRT-PC (Photon Counting mode) and CXO/ACIS-S3 (High Resolution Imaging mode). We set the ACIS-S3 constant value to 1 and calculate an XRT-PC constant of 0.872 using Table~5 from \citet{Plucinsky2017}. For the power-law model with two absorption components, we fix the Galactic contribution to $N_{{\rm{H, MW}}}=1.7\times10^{20}$\,cm$^{-2}$ \citep{Willingale2013} and the redshift to $z = 2.38$ (see Section~\ref{sec:HostGalaxy}). We find no evidence for intrinsic absorption ($N_{\rm H, int}$), so we fix $N_{\rm H, int} = 0$ for our analysis and derive a 3$\sigma$ upper limit of $N_{\rm H, int} < 6\times10^{22}$\,cm$^{-2}$. 

We find evidence for spectral evolution beginning at $\delta t \approx 3.5 \times 10^{-3}$~days. Thus, we split the data into two segments and perform the spectral analysis of each to determine the X-ray photon index, $\Gamma_{\rm X}$. We find $\Gamma_{\rm X, early} = 1.5\pm 0.2$ for $\delta t \lesssim 3.5 \times 10^{-3}$~days, and $\Gamma_{\rm X, late} = 1.8\pm 0.1$ for $\delta t \gtrsim 3.5 \times 10^{-3}$~days. This leads to corresponding spectral indices of $\beta_{\rm X, early} = -0.5 \pm 0.2$ and $\beta_{\rm X, late} = -0.8 \pm 0.1$ ($\beta \equiv 1-\Gamma$, $F_{\nu} \propto \nu^{\beta}$). 

To determine the afterglow unabsorbed fluxes ($0.3-10$\,keV), we apply \texttt{Xspec cflux} only for the afterglow power-law component of the model. The best-fit spectral parameters, counts, and unabsorbed fluxes with $1\sigma$ uncertainties are listed in Table\,\ref{tab:data_xray}. We also derive the $3\sigma$ X-ray flux upper limit from the last CXO observation. We apply the Poissonian confidence levels for small numbers of X-ray events \citep{Gehrels1986} and use the WebPIMMS tool\footnote{\url{https://heasarc.gsfc.nasa.gov/cgi-bin/Tools/w3pimms/w3pimms.pl}} using the late-time spectral parameters (see Table~\ref{tab:data_xray}). We use the inferred $\Gamma_{\rm X}$ to convert the unabsorbed flux to flux density at 1\,keV for subsequent analysis (Table~\ref{tab:data_xray}). 

\subsubsection{Temporal analysis}
\label{sec:Xray_lc_temporal}
The X-ray afterglow light curve exhibits a plateau at $\delta t \lesssim 1.7 \times 10^{-3}$~days followed by a sharp decline with $F_{\rm X} \propto t^{-2.75 \pm 0.09}$ from $ \delta t \lesssim (1.7$--$3.5) \times 10^{-3}$~days (Figure~\ref{fig:xray_image_lc}). This emission is likely not attributed to the external shock \citep{bfc2007RSPTA.365.1213B_Burrows, mgc2010MNRAS.406.2149M_Margutti}, and the corresponding $\Gamma_{\rm X, early}$ may also be unrelated to the shock. Beyond $\delta t \gtrsim 3.5 \times 10^{-3}$~days, most of the data can be fit with a single power-law with $F_{\rm X} \propto t^{-0.70 \pm 0.05}$. However, we note that the X-ray afterglow displays a re-brightening during the second CXO epoch at $\delta t \approx 4.6~$days. To assess the statistical significance, we ran a binomial test with the null hypothesis that the re-brightening is due to a true rise in the count rate, rather than pure statistical fluctuation. We find a $p$-value of $\approx 0.07$ and thus cannot rule out the null hypothesis, indicating a statistically significant deviation from the flux predicted by the single power law for that data point. We revisit this in Section~\ref{sec:reverseshock}. When the re-brightening at $\approx4.6$~days is excluded, the best-fit X-ray decay rate at $\delta t\gtrsim3.5\times10^{-3}$~days is $F_{\rm X}\propto t^{-0.75\pm0.03}$.

\subsection{Optical Upper Limits}

\label{subsec:opt_ag}

\begin{figure*}
    \centering
    \includegraphics[width = \textwidth]{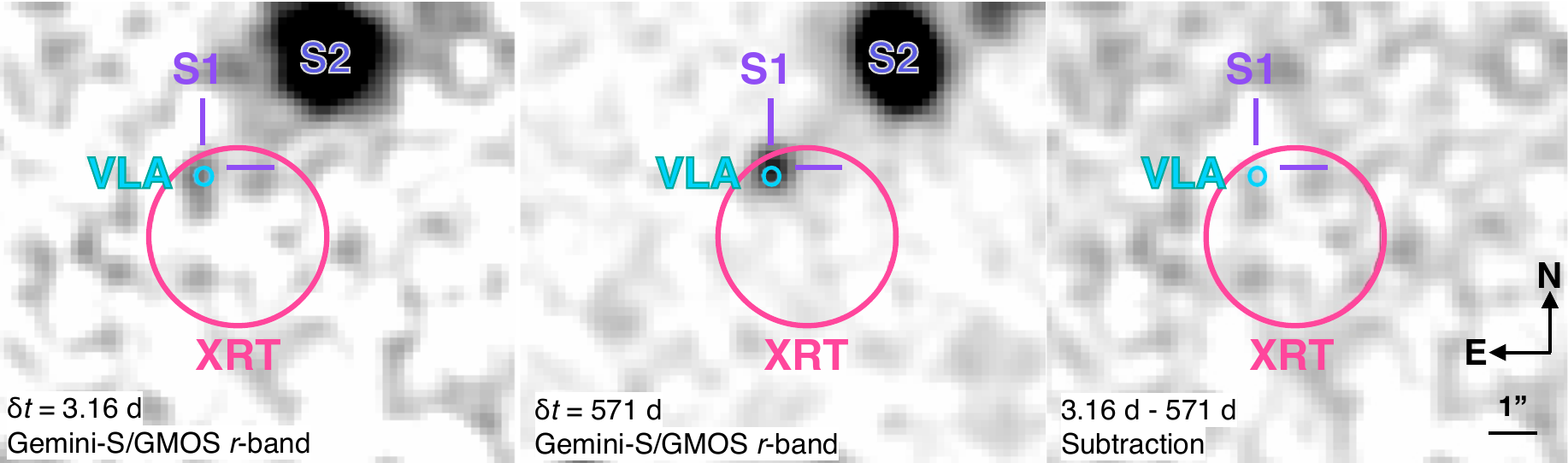}
    \caption{
    \emph{Left:} Gemini-South observations of GRB\,210726A in $r$-band at $\delta t = 3.16$~days post burst. \emph{Middle:} Deep Gemini-South template image of GRB\,210726A in $r$-band at $\delta t = 571$~days post burst. \emph{Right:} Image subtraction between the epoch observed at $\delta t = 3.16$~days and the template image shows no significant residual at the location of the VLA radio source (cyan). We identify ``S1'' (purple crosshairs), a $r = 25$~AB mag source coincident with the radio afterglow, as the host galaxy of GRB\,210726A. No other optical sources are detected within the XRT region (90\% confidence; pink). }
    \label{fig:Host_Galaxy}
\end{figure*}

We imaged the XRT localization region in the $r$-band filter with the Gemini Multi-Object Spectrograph (GMOS) mounted on the 8m Gemini-South telescope (Program GS-2021A-Q-112, PI: Fong) at $\delta t \approx 3.16~{\rm days}$ and $\delta t \approx 5.45~{\rm days}$. We reduce the images using a custom pipeline, {\tt POTPyRI}\footnote{\url{https://github.com/CIERA-Transients/POTPyRI/}}, and register them to SDSS \citep{SDSSDR12} using standard IRAF tasks. A source (``S1'') is marginally detected within the XRT position in the first epoch (Figure~\ref{fig:Host_Galaxy}). We note the presence of an additional source (``S2'', $r = 22.0$~mag) on the Northwest exterior of the XRT position, which was first suggested to be the host galaxy of GRB\,210726A \citep{210726a_GCN30534}.

To assess if the faint source within the XRT region (S1) is afterglow or a coincident galaxy, we obtained a deep $r$-band template image of the field with GMOS on Gemini-South on 2023 February 17 (Program GS-2023A-FT-101, PI: Schroeder). S1 is still clearly detected ($r = 25.06 \pm 0.14$~mag) in the late-time image and is marginally extended. We align the image taken at $\delta t \approx 3.16~{\rm days}$ and the template using standard tasks in IRAF and perform image subtraction between the epochs using \texttt{HOTPANTS} \citep{becker15}. No residual is detected within the XRT region to a limiting magnitude of $r > 24.5$ mag (derived from sources in the field using 
\texttt{IRAF/phot}), indicating that our observation at $\delta t \approx 3.16~{\rm days}$ is minimally contaminated by any optical afterglow. Given the spatial coincidence of S1 with the X-ray (Sections~\ref{sec:Swift_Obs} and \ref{sec:Chandra_Obs}) and radio (Section~\ref{sec:VLA}) afterglows, we identify S1 as the host galaxy. We further discuss the host and associated observations in Section~\ref{sec:HostGalaxy}.

No optical or near-infrared counterpart to GRB\,210726A was reported by the community. We incorporate constraining upper limits from the GCN circulars in our analysis, most notably $r' > 24.4$--$23.0$~mag upper limits at $\approx$2.0-9.5 hours post-burst \citep{2021GCN.30543....1K_CAHA, 2021GCN.30545....1K_OSIRIS/GTC, 210726a_GCN30534}.

\subsection{Radio Afterglow Discovery and Follow-up}

\subsubsection{Very Large Array}
\label{sec:VLA}
\begin{figure}
    \centering
    \includegraphics[width = \columnwidth]{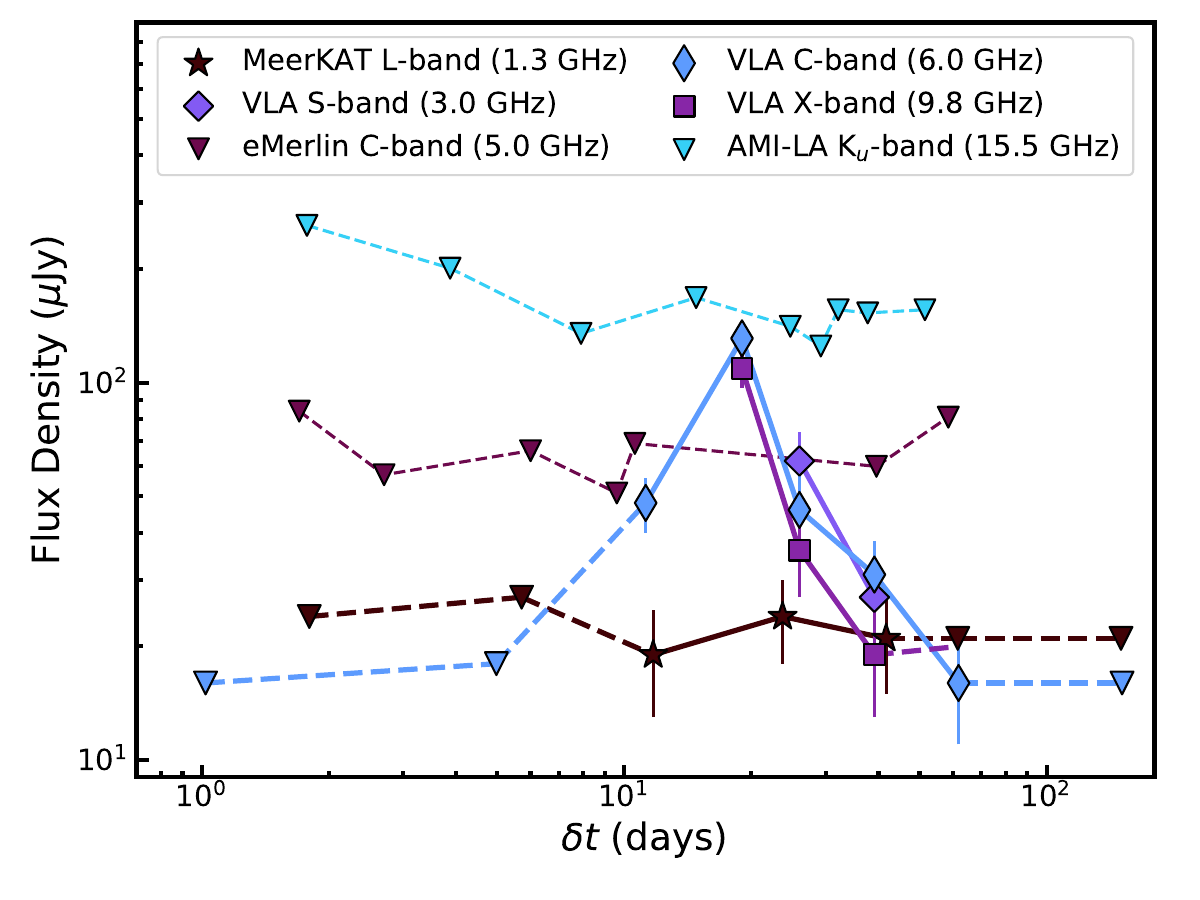}
    \caption{The radio afterglow data for GRB 210726A at 1.3, 3.0, 5.0, 6.0, 9.8 and 15.5\,GHz. Upper limits (3$\sigma$) are denoted using downward-facing triangles; there are only upper limits at 5.0 and 15.5\,GHz. Lines connect observations at the same frequency and are meant to guide the eye. Dashed lines connect non-detections and solid lines connect detections.}
    \label{fig:radio_lc}
\end{figure}

We obtained C-band (4.488-7.612~GHz, central frequency of $6.0~$GHz) observations of GRB\,210726A with the Karl G. Jansky Very Large Array (VLA), at a mid-time of $\delta t \approx 1.0~{\rm days}$ and $\delta t \approx 5.0$~days with Program 20B-057 (PI: Fong). We reduced the data using the Common Astronomy Software Applications Pipeline (\texttt{CASA}, Versions 6.4.1.12, \citealt{CASA}). We used 3C286 for band-pass and flux calibration, and used J1224+2122 for gain and phase calibration in the first and second epochs, respectively\footnote{Upon further inspection, we found the phase calibrators J1224+2122 and J1254+1141 had secondary bright sources resulting in a poor phase solution from the \texttt{CASA} pipeline. To mitigate this issue, we produced a model for each calibrator and its secondary source. We proceeded to manually re-perform the phase calibration steps while incorporating these models. We use a different calibrator in subsequent epochs. \label{phasecal}}. Due to artifacts produced by two bright $\sim$mJy sources within the VLA field of view, we self-calibrated the target field with the prototype automatic self-calibration pipeline developed by the NRAO Science Ready Data Products Initiative\footnote{https://github.com/psheehan/auto\_selfcal}. We do not detect any radio emission in either of the first two epochs, and use the \texttt{pwkit}/\texttt{imtool} program \citep{Pwkit} to measure $3\sigma$ upper limits of $F_{\nu} \lesssim 16\,\mu$Jy and $F_{\nu} \lesssim 18\,\mu$Jy for the first and second epochs, respectively. This supersedes the $3 \sigma$ upper limit we first reported of $F_{\nu}\lesssim 15\,\mu$Jy in \citet{2021GCN.30542....1S_VLAndet} based on a more in-depth treatment of our data here.

We pursued a third C-band observation at a mid-time of $\delta t \approx 11.2$~days, first reported in \citet{2021GCN.30658....1S_VLAdet} using phase calibrator J1254+1141$^{\ref{phasecal}}$, and self-calibrated the target field. We detected a radio source within the XRT error circle, and used \texttt{imtool} to determine $F_{\nu} = 48.3\pm 6.5~\mu{\rm Jy}$. We initiated five additional epochs of multi-frequency observations including C-band, X-band (8.038-11.512 GHz, central frequency of $9.8~$GHz) and S-band, 1.988-4.012~GHz (central frequency of $3.0~$GHz) out to $\delta t \approx 150.84$~days, using both Programs 20B-057 and 21B-198 (PI: Fong). For these five epochs, we used a new phase calibrator, J1327+2210, that had no secondary source issues, and we only self-calibrated the target field. We continue to detect the source at varying fluxes until the final two epochs.

We also use \texttt{imtool} to determine the position of the radio afterglow from the X-band image at $\delta t \approx 19.0$~days. We calculate R.A.=\ra{12}{53}{09.812}, decl.= \dec{+19}{11}{25.09} (J2000) with an uncertainty of $0.12\arcsec$. Due to the variability of the source and the coincidence with the {\emph Swift}/XRT and CXO positions, we consider this to be the radio afterglow of GRB\,210726A. 
To check the fidelity of our flux density measurements of the radio afterglow, we compare the C-band flux densities of four field sources in each of our eight epochs. We find the flux densities of the four sources to be reasonably stable, with variations on the $\sim 10\%$ level. We therefore incorporate a $10\%$ error, added in quadrature to the individual measurement errors, for all flux density measurements.
A summary of these observations can be found in Table~\ref{tab:radio_data}, and the light curves are presented in Figure~\ref{fig:radio_lc}.

\startlongtable
\begin{deluxetable}{ccccc}
 \tabletypesize{\footnotesize}
 \tablecolumns{4}
 \tablecaption{Radio Afterglow Observations}
 \tablehead{ 
    \colhead{Observatory} &
   \colhead{$\delta t^{\rm{a}}$} &
   \colhead{$\nu^{\rm{b}}$} &
   \colhead{Flux density$^{c}$} &\\
      \colhead{} &
   \colhead{(days)} &
   \colhead{(GHz)} &
   \colhead{($\mu$Jy)} &
   }
 \startdata 
VLA 	& $1.0\pm0.04$ 	& 6.0 	& $< 16$ \\
 	& $5.0\pm0.1$ 	& 6.0 	& $< 18$ \\
 	& $11.2\pm0.1$ 	& 6.0 	& $48 \pm 8$ \\
 	& $19.0\pm0.1$ 	& 6.0 	& $131 \pm 14$ \\
 	& $19.0\pm0.1$ 	& 9.8 	& $109 \pm 12$ \\
 	& $26.0\pm0.1$ 	& 3.0 	& $62 \pm 12$ \\
 	& $26.0\pm0.1$ 	& 6.0 	& $46 \pm 11$ \\
 	& $26.0\pm0.1$ 	& 9.8 	& $36 \pm 9$ \\
 	& $39.1\pm0.1$ 	& 3.0 	& $27 \pm 10$ \\
 	& $39.1\pm0.1$ 	& 6.0 	& $31 \pm 7$ \\
 	& $39.1\pm0.1$ 	& 9.8 	& $19 \pm 6$ \\
 	& $61.9\pm0.1$ 	& 6.0 	& $16 \pm 5$ \\
 	& $61.9\pm0.1$ 	& 9.8 	& $< 20$ \\
 	& $150.8\pm0.1$ 	& 6.05 	& $< 16$ \\
\hline
\hline
MeerKAT 	& $1.8\pm0.1$ 	& 1.28 	& $< 24$ \\
 	& $5.7\pm0.2$ 	& 1.28 	& $< 27.0$ \\
 	& $11.7\pm0.2$ 	& 1.28 	& $19 \pm 6$ \\
 	& $23.7\pm0.2$ 	& 1.28 	& $24 \pm 6$ \\
 	& $41.7\pm0.2$ 	& 1.28 	& $21 \pm 6$ \\
 	& $61.6\pm0.1$ 	& 1.28 	& $< 21$ \\
\hline
\hline
eMerlin 	& $1.7\pm0.2$ 	& 5.0 	& $< 84$ \\
 	& $2.7\pm0.2$ 	& 5.0 	& $< 57$ \\
 	& $4.9\pm0.2$ 	& 5.0 	& $< 87$ \\
 	& $6.9\pm0.1$ 	& 5.0 	& $< 69$ \\
 	& $8.0\pm0.3$ 	& 5.0 	& $< 138$ \\
 	& $9.8\pm0.01$ 	& 5.0 	& $< 87$ \\
 	& $10.9\pm0.01$ 	& 5.0 	& $< 69$ \\
 	& $39.5\pm0.2$ 	& 5.0 	& $< 60$ \\
 	& $58.5\pm0.2$ 	& 5.0 	& $< 81$ \\
  \hline
        & $6.0\pm0.3^{d}$ 	& 5.0 	& $< 66$ \\
         & $9.6\pm0.3^{e}$ 	& 5.0 	& $< 51$ \\
    \hline
    \hline
AMI-LA 	& $1.8\pm0.01$ 	& 15.5 	& $< 260$ \\
 	& $3.9\pm0.1$ 	& 15.5 	& $< 200$ \\
 	& $6.8\pm0.1$ 	& 15.5 	& $< 160$ \\
 	& $8.9\pm0.1$ 	& 15.5 	& $< 280$ \\
 	& $14.8\pm0.1$ 	& 15.5 	& $< 170$ \\
 	& $24.7\pm0.1$ 	& 15.5 	& $< 140$ \\
 	& $26.7\pm0.1$ 	& 15.5 	& $< 150$ \\
 	& $28.7\pm0.1$ 	& 15.5 	& $< 120$ \\
 	& $31.6\pm0.1$ 	& 15.5 	& $< 160$ \\
 	& $32.1\pm0.1$ 	& 15.5 	& $< 160$ \\
 	& $34.6\pm0.1$ 	& 15.5 	& $< 200$ \\
 	& $35.7\pm0.1$ 	& 15.5 	& $< 260$ \\
 	& $38.6\pm0.1$ 	& 15.5 	& $< 100$ \\
 	& $45.6\pm0.1$ 	& 15.5 	& $< 520$ \\
 	& $52.6\pm0.01$ 	& 15.5 	& $< 170$ \\
 	& $56.6\pm0.01$ 	& 15.5 	& $< 390$ \\
  \hline
   	& $7.9\pm0.7^{f}$ 	& 15.5 	& $< 140$ \\
        & $29.2\pm1.5^{g}$ 	& 15.5 	& $< 130$ \\
        & $37.7\pm0.7^{h}$ 	& 15.5 	& $< 150$ \\
        & $51.1\pm2.8^{i}$ 	& 15.5 	& $< 160$ \\
  \enddata
\tablecomments{$^{\rm{a}}$ Mid-time of entire observation compared to {\it Swift}/XRT trigger. \\ $^{\rm{b}}$ Central frequency. \\
$^{c}$ Uncertainties correspond to $1\sigma$ confidence. Upper limits correspond to $3 \sigma$. \\
$^{d}$ Combined eMerlin observations at 4.9 and 6.9 days. \\
$^{e}$ Combined eMerlin observations at 8.0, 9.8, and 10.9 days. \\
$^{f}$ Combined AMI-LA observations at 6.8 and 8.9 days. \\
$^{g}$ Combined AMI-LA observations at 26.7, 28.7, and 31.6 days. \\
$^{h}$ Combined AMI-LA observations at 34.6, 35.7, and 38.6 days. \\
$^{i}$ Combined AMI-LA observations at 45.6, 52.6, and 56.6 days. \\
}
\label{tab:radio_data}
\end{deluxetable}

\subsubsection{MeerKAT}

We obtained L-band (0.856-1.711 GHz, mean frequency of 1.28\,GHz, bandwidth of 0.856\,GHz) observations of GRB\,210726A with the MeerKAT radio telescope (in the Karoo desert, South Africa) as part of the ThunderKAT\footnote{The HUNt for Dynamic and Explosive Radio transients with MeerKAT} Large Survey Project \citep[PI: Fender and Woudt,][]{2016mks..confE..13F}. We observed for seven epochs spanning $\approx 1.8$--$61.6~$ days post-burst, with each epoch lasting four hours (with the exception of the first epoch which was 3.4 hours).

We used 3C286 as the flux density and band-pass calibrator and J1330+2509 as the phase calibrator. The data were reduced using a series of semi-automated python scripts designed specifically for producing high-quality images of MeerKAT data \citep[\texttt{oxkat},][]{oxkat}. The \texttt{oxkat} scripts average down the observations to 1024 channels and 8\,s integration times. The data were flagged for radio frequency interference (RFI). A model for 3C286 was derived and applied to the phase calibrator before complex gain calibration using the phase calibrator was performed and applied to the target field, which is then imaged using \texttt{wsclean}. We then used these images as a model for a single round of phase-only self-calibration from which higher dynamic range images were made. 

We detect a source at the VLA position (Section~\ref{sec:VLA}) in the third, fourth and fifth epochs. As a result, GRB 210726A is the first short GRB to be detected by MeerKAT. We measure the flux density using the tool \texttt{imfit} within \texttt{CASA}. We list the flux densities and uncertainties in Table \ref{tab:radio_data} and show the 1.3\,GHz light curve in Figure \ref{fig:radio_lc}. The uncertainties associated with each flux density measurement are calculated by combining the error on the fitting and a 10\% calibration uncertainty in quadrature. 

\subsubsection{\textit{enhanced} Multi-element Radio Linked Interferometer Network}

We searched for a C-band (central frequency 5.01\,GHz, with a 0.51\,GHz bandwidth) counterpart with the \textit{enhanced} Multi-element Radio Linked Interferometer Network (\textit{e}-MERLIN) across nine epochs spanning $\delta t \approx 1.7$--$58.5~$days. We obtained the data through an open-time call proposal (CY12003, PI: Rhodes) and a Director's discretionary time proposal (DD12002, PI: Rhodes). We reduced each epoch using a dedicated \textit{e}-MERLIN pipeline \citep{2021ascl.soft09006M} which uses \texttt{CASA}, \citep[Version 5.6.0, ][]{CASA}, to average down the data into 4\,s integrations and splits the 512\,MHz bandwidth into 512 channels. We flagged the data for RFI and antenna-based flags. We performed band-pass and gain calibration using OQ208 and J1254+1856, respectively, and used 3C286 as the flux calibrator. We applied the calibrator solutions to the target field and imaged the target field using \texttt{tclean}.

We did not detect the afterglow in any of the observations. In order to achieve the deepest limits around the time of the MeerKAT and VLA detections, we concatenated the third and fourth epochs, as well as the fifth, sixth, and seventh epochs, into two deeper observations. We report the $3\sigma$ upper limits of the concatenated epochs, as well as individual upper limits from the seven epochs, in Table~\ref{tab:radio_data}.

\subsubsection{Arcminute Microkelvin Imager - Large Array}

Observations with the Arcminute Microkelvin Imager -- Large Array (AMI--LA) were triggered as a result of the \textit{Swift} alert \citep{Zwart_AMI_2008MNRAS.391.1545Z,  Hickish_AMI_2018MNRAS.475.5677H} and queued up the following day. The first observation commenced at $\delta t \approx 1.7$\,days at K$_u$-band (13.9–17.5 GHz, central frequency of 15.7 GHz). We observed GRB\,210726A for 16 epochs between $\delta t \approx 1.8$--$56.6$~days. We reduced the data using a custom pipeline \texttt{reduce\_dc} \citep{reduce} which flags the raw data for RFI, antenna shadowing, and any effects of poor weather. Amplitude and phase calibration was performed using J1255+1817 followed by flux calibration using 3C286. The calibrated data were exported in \texttt{uvfits} format ready for imaging. 

We flagged the data further and imaged it in \texttt{CASA} (Version 4.7.0) using the tasks \texttt{flagdata} and \texttt{clean}, respectively. After flagging, the effective central frequency of the observations is 15.5~GHz. The afterglow is not detected in any of the epochs. In order to obtain deeper limits, we concatenate the epochs closest in time, also reported in Table~\ref{tab:radio_data}.

\begin{figure*}
    \centering
    \includegraphics[width = 0.9\textwidth]{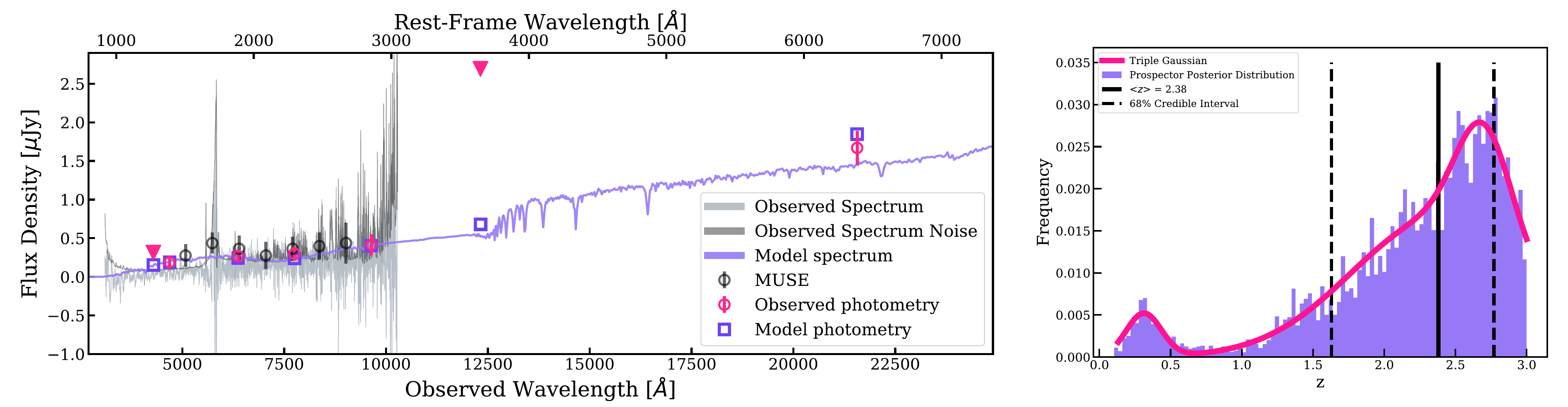}
    \caption{\textit{Left:} The \texttt{Prospector}-derived model spectrum for the host galaxy (purple line) and photometry (purple squares), compared with the observed photometric detections (light pink circles), upper limits (light pink triangles), the photometry from VLT/MUSE for reference (dark grey), and the observed LRIS spectrum (grey line). \textit{Right:} The \texttt{Prospector}-derived redshift posterior distribution (purple histogram), with the median of distribution ($\langle z\rangle=2.38$; solid black line) and the 68\% credible interval (dashed black lines). Given that the redshift distribution favors higher redshift solutions ($z>2$) and the observed photometry is well-fit to the model SED at $z=2.38$, we find that GRB\,210726A likely occurred at $z>2$. {Also shown is a triple Gaussian fit (the sum of three Gaussian distributions, Eq.~\ref{eq:triplegauss}, pink line), that parameterizes the redshift distribution.} }
    \label{fig:prospector}
\end{figure*}
  
\section{Host Galaxy}
\label{sec:HostGalaxy}

\subsection{Host Galaxy Observations}

We obtained additional optical observations of the host galaxy in $griz$-bands with the Gemini Multiple Object Spectrographs (GMOS) mounted on the 8m Gemini-North and Gemini-South telescopes (Program \#GN-2022A-Q-110, PI: Fong; and Program GS-2023A-FT-101, PI: Schroeder, respectively). We also obtained $BR$-band observations with the Low Resolution Imaging Spectrometer (LRIS) mounted on the 10m Keck I (Program \#O300, PI: Blanchard) and $K$-band observations with the Multi-Object Spectrograph (MOSFIRE) on Keck I (Program \#O199, PI: Nugent). Finally, we obtained $J$-band imaging with the MMT and Magellan Infrared Spectrograph (MMIRS) mounted on the 6-m MMT Observatory (Program \#021B-UAO-G177-21B, PI: Nugent). We reduced all images using {\tt POTPyRI} and astrometrically aligned the images to SDSS~DR12 using standard tasks in IRAF. We detect the putative host S1 in all bands except $B$- and $J$-band at a position consistent with those of the X-ray and radio afterglows. We calibrated all images to SDSS \citep{SDSSDR12} (converting to Johnson $BR$ magnitudes for LRIS images using the prescriptions of \citealt{Jester+05}) and performed aperture photometry with IRAF/\texttt{phot} using seeing-based apertures.

We calculate the probabilities of chance coincidence (P$_{\rm cc}$; Figure~\ref{fig:Host_Galaxy}) for S1 and S2 \citep{Bloom+02} using their offsets from the VLA position and $r$-band magnitudes from our Gemini-S imaging. We determine P$_{\rm cc, S1} = 5 \times 10^{-4}$ and P$_{\rm cc, S2} = 0.040$ (see also \citealt{BRIGHT_I}). We furthermore find no host candidate at a larger angular distance in our deep imaging. We thus conclude that S1 is the host galaxy of GRB\,210726A. We list the optical photometry of the host galaxy in Table~\ref{tab:hostphot}, and show the position of the afterglow in relation to the host galaxy in Figure~\ref{fig:Host_Galaxy}.

\begin{deluxetable}{cccc}
\tabletypesize{\footnotesize}
\label{tab:hostphot}
\centering
\tablecolumns{9}
\tabcolsep0.06in
\tablecaption{Host Galaxy Photometry}
\tablehead {
\colhead {Telescope}	&
\colhead {Instrument}		&
\colhead {Filter}	&
\colhead {Magnitude (AB)}
}
\startdata
Keck I & LRIS & $B$ & $> 26.1$ \\
Gemini-N & GMOS & $g$ & $25.8 \pm 0.3$ \\
Keck I & LRIS & $R$ & $25.4 \pm 0.2$ \\
Gemini-S & GMOS & $r$ & $25.1 \pm 0.1$ \\
Gemini-N & GMOS & $i$ & $25.1 \pm 0.2$ \\
Gemini-N & GMOS & $z$ & $24.8 \pm 0.2$  \\
MMT & MMIRS & $J$ & $> 22.9$  \\
Keck & MOSFIRE & $K$ & $23.3 \pm 0.1$  \\
\hline
\hline
Telescope & Instrument & Wavelength Bin (\AA) & Magnitude (AB)  \\
\hline
VLT & MUSE & 4750--5407 & $25.3\pm0.6$ \\
VLT & MUSE & 5407--6064 & $24.8\pm0.4$ \\
VLT & MUSE & 6065--6722 & $25.0\pm0.5$ \\
VLT & MUSE & 6722--7379 & $25.3\pm0.7$ \\
VLT & MUSE & 7379--8036 & $25.0\pm0.5$ \\
VLT & MUSE & 8036--8693 & $24.9\pm0.5$ \\
VLT & MUSE & 8694--9351 & $24.8\pm0.7$ \enddata
\tablecomments{All photometry is corrected for Milky Way extinction in the direction of the burst according to \citep{schlafley2011ApJ...737..103S}.}
\end{deluxetable}

We observed the field of GRB 210726A with the Multi Unit Spectroscopic Explorer (MUSE) at the Very Large Telescope (VLT) on 28 April 2023 with 4$\times$247 s exposures on target (Program ID 110.24CF, PI: Tanvir). The MUSE nominal wide-field mode has a spatial field of view of 1 arcmin on the side and covers a wavelength range of 4750--9350 {\AA}. The data were processed with the EsoRex MUSE pipeline version 2.8.1 \citep{weilbacher2020}. Besides the standard processing steps for spectroscopy, the pipeline includes flux-calibration and correction for telluric absorption lines. The seeing measured in the final stacked data cube is $\sim1\arcsec$. A white light image created by co-adding the integral field spectroscopy data cube along the spectral direction reveals a faint source consistent with the position of the GRB host galaxy, S1. We extract a one-dimensional spectrum and its associated error spectrum within a 1 arcsec diameter aperture centered on S1. The extracted spectrum has a low signal-to-noise ratio and is dominated by residuals from telluric lines particularly redwards of 7800 {\AA}. We do not detect any significant emission lines in the spectrum, nor do we clearly identify a 4000 {\AA} break, so a definite spectroscopic redshift cannot be determined. To compare to the photometry and host galaxy modeling fits, we split the MUSE spectrum into seven identically sized wavelength bins, and extract the integrated flux of the host galaxy\footnote{\label{Muse}The sky subtraction errors can in some cases enhance the flux level artificially at these faint $2 \sigma$ detection levels} as presented in Table~\ref{tab:hostphot}.

On 28 April 2023 UT, we obtained $4\times1200$~s of spectroscopy of S1 with Keck/LRIS (Program \#O199, PI: Nugent). The spectrum covers the optical wavelengths of $\sim$3200--10200~\AA. We used the Python Spectroscopic Data Reduction Pipeline (\texttt{PyPeIt}; \citealt{phw+2020}) to reduce and extract the spectrum. With \texttt{PypeIt}, we applied an overscan subraction, flat field correction, wavelength calibration, and combined all 2D frames using the \texttt{PyPeIt} built-in function. We extracted the 1D trace and variance, which we converted to an error spectrum, from the coadded image using the \texttt{boxcar} method with a $1.5\arcsec$ radius to encompass the full trace and possible spectral features. We applied a flux calibration using the spectrophotometric standard star BD284211, taken on the same night. We detect a faint continuum in both the 1D and 2D images which surpasses the noise level at $\approx 5000~$\AA, however we see no evidence for even low signal-to-noise spectral lines that could confirm the redshift of the host.

\subsection{Host Galaxy Modeling}
\label{sec:hostgalaxymodel}
We use the observed photometric upper limit and detections (Table~\ref{tab:hostphot}),  in combination with the stellar population modeling code \texttt{Prospector} \citep{Leja_2017}, to attempt to constrain the redshift of the host galaxy. \texttt{Prospector} applies a nested sampling fitting routine \citep{speagle2020} to the observed data to create posterior distributions of the stellar population properties of interest, as well as generate model SEDs \citep{Conroy2009, Conroy2010}. We employ a Chabrier initial mass function (IMF) \citep{Chabrier2003}, a parametric delayed-tau star formation history (SFH), the Milky Way dust extinction law \citep{Cardelli1989}, and allow redshift to be a sampled parameter\footnote{{This method of determining a photometric redshift for a short GRB host is common practice in the absence of a spectroscopically determined redshift \citep[e.g.][]{2022MNRAS.515.4890O, BRIGHT_II}.}} ranging $0.1 < z < 3.0$. See \cite{BRIGHT_II} for further details on the stellar population modeling. We find that the median \texttt{Prospector}-derived redshift is $z = 2.38^{+0.39}_{-0.75}$. We show the model SED derived at this redshift and the redshift posterior distribution in Figure \ref{fig:prospector}. We note that the redshift posterior distribution is double peaked, however with much higher probability lying towards a high redshift ($z>2$) solution. We furthermore compare the ten photometric points from the VLT/MUSE data. The MUSE data$^{\ref{Muse}}$ are consistent with the observed photometry used in the \texttt{Prospector} fit, and show no indication of continuum features that would lead to a more conclusive redshift determination. Given that the observed optical photometry and Keck/LRIS spectrum are featureless and show no indication of a 4000\AA\ break, we infer that the break must occur after the $z$-band detection, making the redshift of this host likely at $z\gtrsim 1.2$. We use the median redshift $z=2.38$ for our afterglow modeling (Section~\ref{sec:fs_model}).

\section{Basic properties of the afterglow}\label{sec:afterglow}

To understand the multi-wavelength behaviour of GRB\,210726A, we consider the X-ray to radio afterglow in the framework of synchrotron emission produced by electrons accelerated into a non-thermal power law distribution ($N(\gamma_{\rm e}) \propto \gamma_{\rm e}^{-p}$, with expected values of $2<p<3$) by the forward shock (FS) of the GRB (e.g. \citealt{spn1998_sari, Wijers1999ApJ...523..177WG1999, PanaitescuKumar2000ApJ...543...66P, GS2002}). This emission is characterized by three break frequencies: the self-absorption frequency, $\nu_{\rm sa}$, the characteristic frequency, $\nu_{\rm m}$, and the cooling frequency, $\nu_{\rm c}$, which are connected by a broken power law spectrum normalized by the characteristic flux, $F_{\nu, \rm m}$ \citep{GS2002}. Throughout this analysis, we use the convention $F_{\nu} \propto t^{\alpha} \nu^{\beta}$, where $\alpha$ and $\beta$ are the temporal and spectral power-law indices respectively.

Additionally, we consider the effects of collimation. The standard FS synchrotron model assumes a spherically symmetric blast wave, which is valid as long as the angular size of the beaming angle ($\theta_{\rm beam} = 1/\Gamma$, where $\Gamma$ is the Lorentz factor of the jet) is less than the true opening angle of the jet ($\theta_{\rm jet}$). Once the size of $\theta_{\rm beam}$ approaches $\theta_{\rm jet}$, the observed light curves steepen achromatically (``jet break") which occurs at time $t_{\rm jet}$ \citep{Rhoads1999, SariPiranHalpern1999}. The afterglow after the jet break is expected to fall as $F_{\nu} \propto t^{-p}$ when $\nu_{\rm obs} > \nu_{\rm m}$ (assuming lateral spreading, where $\nu_{\rm obs}$ is the observing frequency).

\subsection{The location of the break frequencies}
\label{sec:breakfreq}

\begin{figure}
    \centering
    \includegraphics[width = \columnwidth]{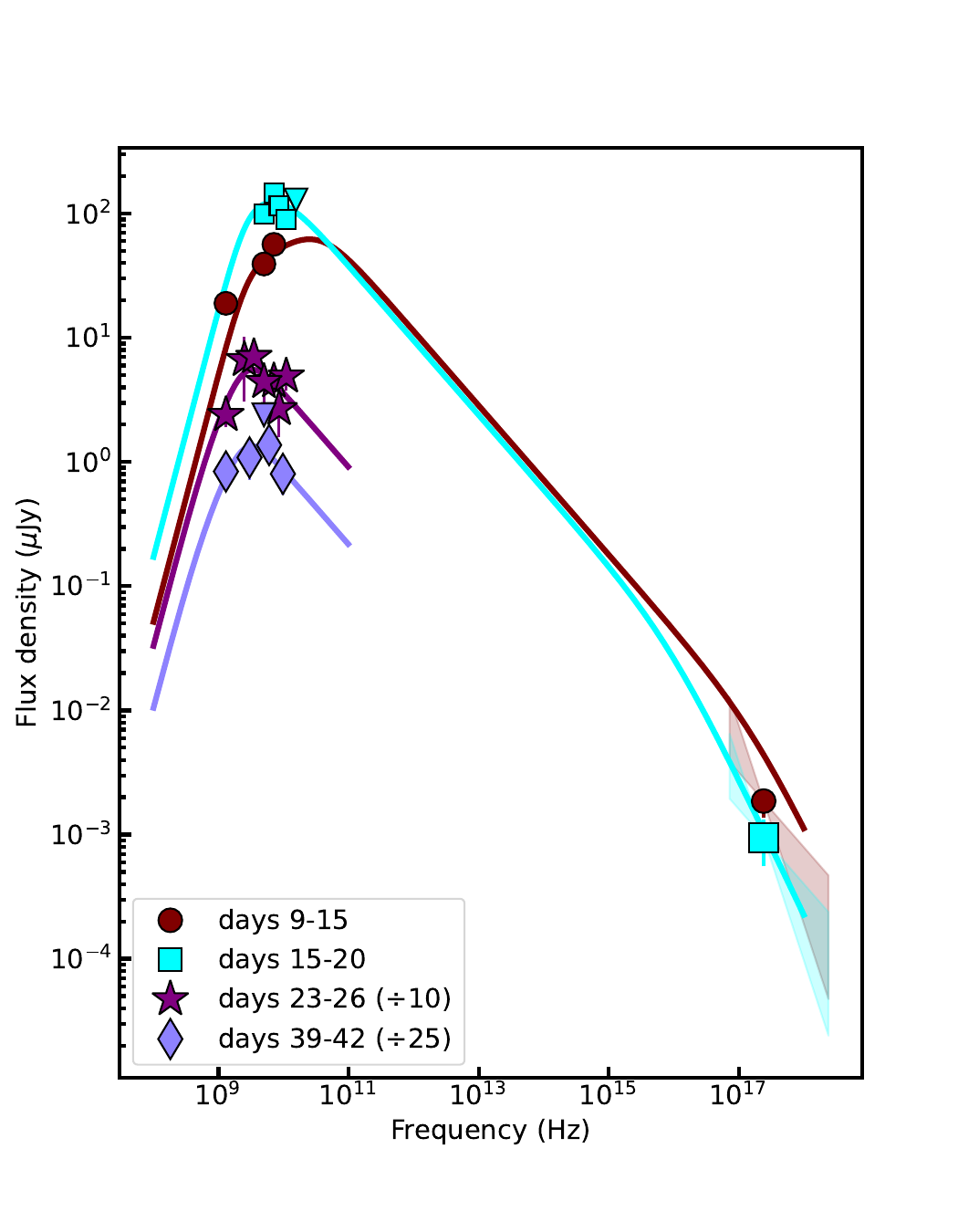}
    \caption{{The radio to X-ray afterglow spectral energy distribution for four time bins ($\delta t\approx $\,9--15, 15--20, 23--29 and 39--42\,days). Triangles represent $3 \sigma$ upper limits, all other symbols represent detections}. {Lines represent s}pectral fits to the radio and X-ray data where the spectra consist of four power laws split by three frequency breaks: $\nu_{\rm sa}$, $\nu_{\rm m}$ and $\nu_{\rm c}$. The four power laws follow: $\nu^{2}$, $\nu^{1/3}$, $\nu^{-0.6}$ and $\nu^{-1.1}$, where the latter two are derived from $p = 2.2$.}
    \label{fig:joint_fit}
\end{figure}

The radio and X-ray afterglows allow us to place constraints on the locations of $\nu_{\rm sa}$, $\nu_{\rm m}$, $\nu_{\rm c}$, and $F_{\nu, \rm m}$, as well as their temporal evolution. Here, we split the VLA detections into lower and upper sidebands where SNR permits (Table~\ref{tab:VLA_epoch3_4_5_frequency}). We group the data into four epochs: $\delta t\approx $\,9--15, 15--20, 23--29 and 39--42\,days (Figure \ref{fig:joint_fit}); the first two are accompanied by CXO observations. 

We first investigate the SEDs of the $\delta t \approx 9$--$15$~days observations\footnote{While the $\Delta t/t$ for the $\delta t \approx 9$--$15$~days SED is large, the data at $\approx 9$~days is a non-detection from \textit{e}-Merlin, and the data at $\approx 15$~days is from the X-ray afterglow, which has settled into a power law decay. All of the radio detections in this period are at $\approx 11$~days.}. The radio spectrum can be fit with a single power law, $\beta = 0.6 \pm 0.1$. This is too shallow for optically thick synchrotron emission ($\nu^{5/2}$ or $\nu^{2}$) but too steep for $\nu_{\rm sa} < \nu_{\rm obs} < \nu_{\rm m}$ ($\nu^{1/3}$), indicating that $1.3~{\rm GHz} < \nu_{\rm sa} < 6.0~{\rm GHz} < \nu_{\rm m}$.  Additionally, the negative spectral index between the 6~GHz afterglow and the X-rays ($\beta_{\rm 6GHz~to~X-ray} = -0.6 \pm 0.02$) demonstrates that $\nu_{\rm m}$ must fall between them. Therefore, we find at least two spectral breaks encompassed by our data.

To further constrain the locations of the break frequencies, we fit the $\delta t \approx 9$--$15$~days epoch with a triple broken power law, where we fix the spectral indices to $\beta_{1} =2$, $\beta_{2} = 1/3$, $\beta_{3} = (1-p)/2$ and $\beta_{4} = -p/2$, with a smoothing parameter of $s = 2$. We fix $\beta_{3} = -0.6$, from our calculated $\beta_{\rm 6GHz~to~X-ray}$ (which corresponds to $p = 2.2$) and $\beta_{4} = -1.1$. This fit results in $F_{\nu, \rm m} \approx 60~\mu$Jy, $\nu_{\rm m} \approx 20~$GHz, and indicates $\nu_{\rm c} \sim \nu_{\rm X}$ (where $\nu_{\rm X}$ is the central frequency of the X-ray band) at this time. Furthermore, the fit results in $\nu_{\rm sa} \approx 2~$GHz, and as this break frequency does not evolve in an ISM environment \citep{GS2002}, we keep $\nu_{\rm sa}$ stationary in our subsequent fits.

At $\delta t \approx 15$--$20~$days, the radio detections are over a smaller frequency range and the radio SED is relatively constant, with flux density values of $F_{\nu} \approx 100$--$150~\mu$Jy for $\nu \approx 5$--$11~$GHz. We similarly fit this epoch with a triple broken power law, and find $\nu_{\rm m}\sim 8$\,GHz, and $F_{\nu, \rm m} \approx 120~\mu$Jy. Such an increase in $F_{\nu, \rm m}$ is not expected in the standard FS scenario ($F_{\nu, \rm m} =~$constant, prior to jet break, \citealt{GS2002}), and we return to this in Section~\ref{sec:RadioFlare}.  
We also find that at this time, $\nu_{\rm c} < \nu_{\rm X}$, indicating $\nu_{\rm c}$ is evolving to lower frequencies, as expected for the FS model in an ISM environment. For $p = 2.2$ and $\nu_{\rm c} < \nu_{\rm X}$, we expect $\alpha_{\rm X} = -1.2$ and $\beta_{\rm X} = -1.1$, both of which are steeper than the measured values of $\alpha_{\rm X} = -0.75 \pm 0.03$ and $\beta_{\rm X} = -0.8 \pm 0.1$ (Sections~\ref{sec:Xray_lc_spectra} \& \ref{sec:Xray_lc_temporal}). However, we note that processes like inverse Compton (IC) cooling \citep{SariEsin2001, lbm+2015ApJ...814....1L} and Klein-Nishina (KN) corrections \citep{nas_2009ApJ...703..675N_nakar, JBv2021MNRAS.504..528J_Jacovich} lead to shallower spectral and temporal indices at high energies, which could explain this discrepancy. We return to this point in Section~\ref{sec:MCMC_fit_to_all}. The radio SEDs at $\delta t \approx 23$--$29$ and $39$--$42~$days are best described by a broken power law. We find that they are both consistent with $\beta_{1} \approx 2$ and $\beta_{2} \approx -0.6$, indicating $\nu_{\rm m} \approx \nu_{\rm sa}$ during these epochs \citep{GS2002}.

In summary, we expect $p \approx 2.2$, $\nu_{\rm sa} \approx 2~$GHz, $\nu_{\rm m}$ to be evolving through the radio band, and $\nu_{\rm c} \lesssim \nu_{\rm X}$.

\subsection{A radio flare at 6~GHz}
\label{sec:RadioFlare}

Given the wealth of radio observations, we next investigate the behavior of the radio light curves of GRB\,210726A. There is a rapid rise in flux observed in the C-band light curve ($\alpha = 1.9 \pm 0.3$) from $\delta t \approx 11.2$--$19.0~$days, as well as an apparent rise in $F_{\nu, \rm m}$, based on our triple broken power law fit (Section~\ref{sec:breakfreq}). Furthermore, at $\delta t \gtrsim 19.0~$days, all of the radio light curves except L-band display a rapid decline, with the C-band light curve falling as $\alpha = -2.1 \pm 0.4$ (Figure~\ref{fig:radio_lc}).  

The rise of the C-band light curve is not consistent with standard afterglow models, and given its subsequent steep decline, we refer to this feature as the ``radio flare". While we cannot readily explain the rise of the C-band light curve with standard FS models (Section~\ref{sec:fs_model}), the apparent achromatic decline of the radio light curves may be attributed to the passage of $\nu_{\rm m}$, a jet break, or a combination of the two. With the cadence of our observations, it is not possible to break the degeneracy of these effects with this simple analysis, and we therefore place limits on a jet break occurring at $\gtrsim 19~$days.

\section{Forward shock modelling}\label{sec:fs_model}

\subsection{Model Description}

We fit the radio to X-ray afterglow data of GRB\,210726A using the modeling framework laid out in \citet{lbt2014ApJ...781....1L_laskar} to derive the burst properties. In addition to the standard FS framework, our modeling incorporates the scattering effects of scintillation \citep{gn2006ApJ...636..510G}, 
IC cooling \citep{SariEsin2001, lbm+2015ApJ...814....1L}, and KN corrections \citep{nas_2009ApJ...703..675N_nakar, JBv2021MNRAS.504..528J_Jacovich}.

The light curves and SEDs of the afterglow are parameterized by: $p$, the isotropic equivalent kinetic energy $E_{\rm K, iso}$, the circumburst density profile $\rho = A r^{k}$ (where  $k = 0$ and $A = m_{\rm p} n_{0}$ for an ISM environment), and the fraction of energy imparted to the electrons, $\epsilon_{\rm e}$, and magnetic field, $\epsilon_{\rm B}$. We further require that $\epsilon_{\rm e} + \epsilon_{\rm B} < 1$, and assume that the fraction of participating, non-thermal (NT) electrons ($f_{\rm NT}$) is 1 \citep{ew2005}. Additionally, \citet{2023MNRAS.523..775G} use the optical non-detection at $\delta t \approx 0.09~$days to classify GRB\,210726A as a ``dark" GRB, in which the optical afterglow is suppressed by dust along the line of sight of the GRB. To account for this, we include dust extinction, $A_V$ \citep[assuming an SMC extinction law,][]{Pei1992ApJ...395..130P}, as a free parameter. We fix the redshift to $z = 2.38$ (see Section~\ref{sec:hostgalaxymodel}).

\begin{figure}
    \centering
    \includegraphics[width = 0.48\textwidth]{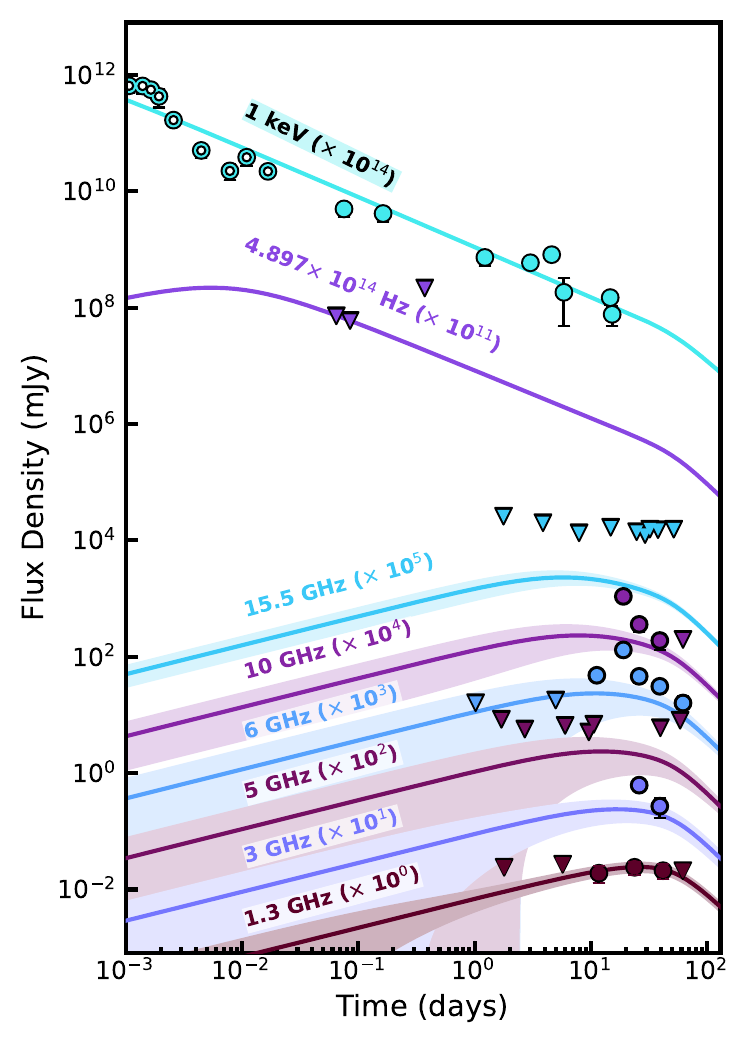}
    \caption{Forward shock (FS) fit of the afterglow of GRB\,210726A. Circles indicate detections, where as triangles indicate $3 \sigma$ limits. Open circles indicate data not included in the fit. Shaded regions indicate predicted variability due to scintillation. It is clear that the FS model alone cannot adequately explain the radio light curves, thus revealing a radio flare. }
    \label{fig:forwardshockfits}
\end{figure}

\begin{deluxetable*}{c|cc|cc}
 \tabletypesize{\footnotesize}
 \tablecolumns{5}
 \tablecaption{Forward Shock Parameters}
 \tablehead{  
    \colhead{} &
    \twocolhead{Redshift Fixed} & 
    \twocolhead{Redshift Free}
   }
 \startdata 
 Parameter & Best Fit Model & MCMC Results & Best Fit Model & MCMC Results \\
 \hline
 $z$ & $2.38$ & -- & $2.74$ & $2.29^{+0.44}_{-0.63}$\\
$p$ & $2.04$ & $2.05^{+0.03}_{-0.02}$ & $2.03$ & $2.06^{+0.03}_{-0.02}$ \\
$E_{\rm K, iso}~(\times 10^{52}~{\rm erg})$ & $8.1$ & $3.8^{+2.6}_{-2.0}$ & $11.6$ & $3.1^{+3.2}_{-1.9}$ \\
$n_0~({\rm cm}^{-3})$ & $7.4\times 10^{-2}$ & $5.9^{+5.2}_{-3.1} \times 10^{-2}$ & $5.8\times 10^{-2}$ & $6.6^{+5.6}_{-3.3} \times 10^{-2}$ \\
$\epsilon_{\rm e}$ & $9.0\times 10^{-1}$ & $6.1^{+2.3}_{-2.0} \times 10^{-1}$ & $9.7\times 10^{-1}$ & $6.4^{+2.2}_{-2.1} \times 10^{-1}$ \\
$\epsilon_{\rm B}$ & $1.1\times 10^{-4}$ & $5.1^{+33.3}_{-3.6} \times 10^{-4}$ & $1.1\times 10^{-4}$ & $4.7^{+29.6}_{-3.3} \times 10^{-4}$ \\
$t_{\rm jet}~({\rm days})$ & $51.0$ & $55.5^{+40.8}_{-19.9}$ & $49.3$ & $58.6^{+51.2}_{-21.3}$ \\
$\theta_{\rm jet}~({\rm deg})$ & $14.7$ & $16.2^{+3.3}_{-2.3}$ & $13.4$ & $17.2^{+4.7}_{-2.9}$ \\
$E_{\rm K}~(\times 10^{52}~{\rm erg})$ & $2.6\times 10^{-1}$ & $1.5^{+1.2}_{-0.8} \times 10^{-1}$ & $3.2\times 10^{-1}$ & $1.4^{+1.4}_{-0.8} \times 10^{-1}$ \\
$A_{V}~(\rm mag)$ & $\gtrsim 0.72$ & -- & $\gtrsim 0.86$ & --  \\
\hline
\multicolumn{5}{c}{Break Frequencies and Peak Flux at $\delta t = 11.2$~days} \\
\hline
$\nu_{\rm sa}~$(Hz) & $1.5 \times 10^{8}$ & -- & $1.6 \times 10^{8}$ & -- \\
$\nu_{\rm m}~$(Hz) & $9.5 \times 10^{9}$ & -- & $9.8 \times 10^{9}$ & --\\
$\nu_{\rm c}~$(Hz) & $1.5 \times 10^{15}$ & -- & $6.4 \times 10^{15}$ & --\\
$F_{\nu, \rm m}~$($\mu$Jy) & $47$ & -- & $53$ & --\\
\enddata
\tablecomments{\emph{Top:} The best-fit
and summary statistics (median and 68\% credible intervals) parameters from the marginalized posterior density functions of the forward shock (FS) afterglow parameters from our MCMC modeling for the redshift fixed (left, $z = 2.38$) and redshift free (right) afterglow modeling fits. The parameters of the best-fit model may differ from the summary statistics as the former is the peak of the likelihood distribution and the latter is calculated from the full marginalized posterior density functions of each parameter.
\emph{Bottom:} The break frequencies and peak flux of the FS from the best fit parameters at $\delta t = 11.2~$days for the redshift fixed (left) and redshift free (right) fits.}
\label{tab:FS_fit}
\end{deluxetable*}
\subsection{MCMC Modeling}
\label{sec:MCMC_fit_to_all}

Using the Markov Chain Monte Carlo Python-based code \texttt{emcee} \citep{emcee_2013PASP..125..306F}, we use 128 walkers and 10k steps, discarding the first $\sim 0.4\%$ steps as burn-in as the likelihood of the model has not reached a stable value. Our MCMC fit results in a $\chi^2/d.o.f. \approx 69/24 \approx 2.9$ and maximum log-likelihood ($\mathcal{L}$) of $\approx 117$. The ordering of the break frequencies is $\nu_{\rm sa} < \nu_{\rm m} < \nu_{\rm c}$ for the entire duration of the detected afterglow. We find $p \approx 2.04$, similar to our expectations from the broken power law fits to the radio to X-ray SEDs (Section~\ref{sec:breakfreq}), and the parameters of the highest likelihood model are $E_{\rm K,iso} \approx 8.1 \times 10^{52}~$erg, $n_{\rm 0} \approx 7.4 \times 10^{-2}~{\rm cm}^{-3}$, $\epsilon_{\rm e} \approx 0.90$ and $\epsilon_{\rm B} \approx 1.1 \times 10^{-4}$. The derived value of $\epsilon_{\rm e}$ is higher than the expected equipartion value of $1/3$. However, this tension can be alleviated if $f_{\rm NT} < 1$, and we discuss this further in Appendix~\ref{sec:MeerKAT_masked}.

We also find $\nu_{\rm c} < \nu_{\rm X}$, as expected. IC cooling and KN corrections become important in the regime when $\epsilon_{\rm B} \ll \epsilon_{\rm e}$ \citep{PanaitescuKumar2000ApJ...543...66P, SariEsin2001, GS2002, nas_2009ApJ...703..675N_nakar, JBv2021MNRAS.504..528J_Jacovich}. For these parameters, we find $\nu_{\rm m} < \hat{\nu}_{\rm c}<\nu_{\rm c}<\nu_{\rm X}<\hat{\nu}_{\rm m},\nu_0$, where the $\hat{\nu}$ are additional breaks in the synchrotron spectrum arising from KN corrections and $\nu_0$ corresponds to the frequency above which IC cooling is strongly KN suppressed and therefore no longer important \citep{nas_2009ApJ...703..675N_nakar}. In this regime, the predicted X-ray spectral index is $\beta_{\rm X}=3(1-p)/4\approx-0.78$, consistent with the value of $\beta_{\rm X,late}=-0.8\pm0.1$ from the \textit{Swift}/XRT analysis (Section~\ref{sec:Xray_lc_spectra}).

{From the derived values of $E_{\rm K, iso}$, $n_0$, and $t_{\rm jet}$, t}he resulting value of $\theta_{\rm jet} \approx 14.7^\circ$ is wide compared to most short GRBs with measured $\theta_{\rm jet}$ values \citep{refb+2022}. We also find that $A_{V} \gtrsim 0.72~$mag is required in order to not violate the optical non-detections at $\delta t \approx 0.08$--$0.09~$days, consistent with dust as the cause of optical darkness and similar to the value inferred host median value \citep{BRIGHT_II}. We present the full X-ray to radio afterglow model in Figure~\ref{fig:forwardshockfits}, and list the best fit values and summary statistics (medians with 68\% credible intervals) of all fit parameters in Table~\ref{tab:FS_fit} 
{(left, ``redshift fixed'')}. 

The L-band light curve of this model is well fit (Figure~\ref{fig:forwardshockfits}), with the rise and fall corresponding to the passage of $\nu_{\rm m}$. This is at odds with expectations from the preliminary spectral analysis (Section~\ref{sec:breakfreq}), where we expected $\nu_{\rm sa} > 1.4$~GHz. In fact, in this fit, $\nu_{\rm sa} \approx 0.2$~GHz, an order of magnitude lower than predicted. We explore this further Appendix~\ref{sec:MeerKAT_masked}. Additionally, much of the S-, C-, and X-band light curves are under-predicted, with the C-band model light curve discrepant by a factor of $\approx 1.7$--$6.0$ (ignoring the radio flare at $\approx 19~$days brings this range down to $\approx 1.7$--$2.3$, Figure~\ref{fig:forwardshockfits}). 

{To explore the dependence of the afterglow parameters on redshift, we also perform an MCMC fit with $z$ as a free parameter. We utilize the redshift posterior derived by the \texttt{Prospector} modeling (Section~\ref{sec:hostgalaxymodel}) as a prior. To do so, we parameterize the \texttt{Prospector} redshift posterior using a triple Gaussian fit (the sum of three Gaussian distributions), such that the prior function is:}
\begin{multline}
    P(z) = \frac{1}{\sqrt{2 \pi}} \sum_{i=1}^{3} \frac{c_i}{\sigma_i} \exp\left[{-\frac{1}{2} \left( \frac{z-\mu_{i}}{\sigma_i}\right)^2}\right]; \\
    \quad z \in [0,3]
    \label{eq:triplegauss}
\end{multline}
{where $\mu_{i}$, $\sigma_{i}$ are the mean and standard distribution (respectively) of each Gaussian, and $c_i$ is a normalization factor. We find $\mu_1,~\mu_2,~\mu_3 \approx 0.31,~2.39,~2.70$, $\sigma_1,~\sigma_2,~\sigma_3 \approx 0.12,~0.62,~0.19$, and $c_1,~c_2,~c_3 \approx 1.6 \times 10^{-3},~2.6 \times 10^{-2},~6.3 \times 10^{-3}$ parameterize the \texttt{Prospector} posterior well (Figure~\ref{fig:prospector}, right). We then normalize $P(z)$, such that $\int_0^3 P(z) \,{\rm d}z = 1$, as required by our afterglow fitting code. The summary of the resultant MCMC fit with redshift as a free parameter can be found in Table~\ref{tab:FS_fit} (right, ``redshift free''). Due to the high value of $P(z)$ at $z \gtrsim 2$, the highest likelihood afterglow fit settles on a redshift of $z \approx 2.74$. The afterglow parameters from the redshift free fit are similar to our afterglow fit with the redshift fixed to $z = 2.38$. We therefore focus our discussion with the parameters from the $z = 2.38$ afterglow fit.}

To explore the strength of the constraints imposed on the model by the L-band observations, and to investigate whether it is possible to improve the match to the high-frequency data, we also attempt a fit after masking the MeerKAT data. The resultant fit over-predicts the L-band observations by a factor of $\approx1.5$ and still does not significantly better match the higher frequency radio observations, and is therefore disfavored. We present this model for completeness in Appendix~\ref{sec:MeerKAT_masked}.

\section{Additional Components to Explain the Radio Flare}
\label{sec:Alternative_theories}

A single FS does not adequately describe the broadband afterglow, and in particular the emission at 6\,GHz, of GRB\,210726A. Here we explore two effects to explain the radio flare: an energy injection period into the FS and an RS.  Interstellar scintillation, an off-axis afterglow, and variable external density scenarios are presented in Appendix~\ref{sec:extrinsicscenarios}, but are ultimately disfavored. 

\begin{figure}
    \centering
    \includegraphics[width = 0.48\textwidth]{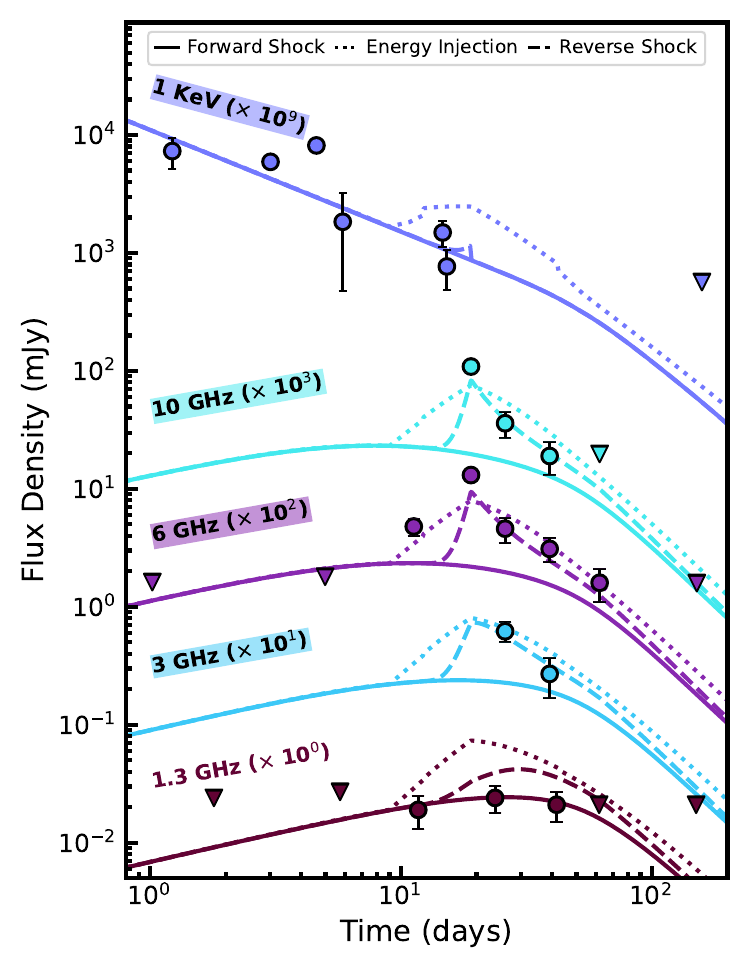}
    \caption{The radio and X-ray afterglow of GRB\,210726A and the forward shock (FS) model presented (solid line). The dotted lines correspond to the FS fit with an additional energy injection period. The dashed lines correspond to the FS fit with an additional reverse shock (RS) component.}
    \label{fig:FlarePlot}
\end{figure}

\subsection{Energy Injection}
\label{sec:energyinjection}

Our first scenario uses energy injection to explain the radio flare observed at C-band. There are two distinct models: the transfer of braking radiation from a millisecond magnetar into the FS \citep{ Dai1998, 2000ApJ...535L..33S, 2001ApJ...552L..35Z} and a stratified jet with a distribution of ejecta Lorentz factors, where the ejecta shells are released over a short period \citep[short compared to the timescale of the afterglow,][]{Rees1998, 2000ApJ...535L..33S}. In both cases, energy injection into the FS is only observable as a rapid brightening if the injection is rapid and the injected energy dominates over the energy of the initial blast wave caused by the fastest moving ejecta \citep{lbm+2015ApJ...814....1L}. Before the period of injection, the light curves follow a standard FS model with a constant kinetic energy. The energy injection results in an achromatic re-brightening of the afterglow. After the end of energy injection, the evolution returns to that of a blast wave with increased kinetic energy. 

In the case of a millisecond magnetar central engine, the injected energy is directly related to the luminosity function index of the central engine, $q$, such that the luminosity follows $L \propto t^{-q}$, $E_{\rm K, iso} \propto t^{1-q}$ \citep{Dai1998, 2001ApJ...552L..35Z}. In the stratified ejecta scenario, the injected energy is related to the kinetic energy of the ejecta mass, where some mass, $M$, is moving with Lorentz factors greater than the bulk Lorentz factor $\Gamma$ (i.e. $M(> \Gamma) \propto \Gamma^{-s}$), resulting in $E_{\rm K, iso} \propto \Gamma^{1-s}$ \citep{Rees1998, pmr1998ApJ...503..314P, 2000ApJ...535L..33S}. Based on the afterglow alone, it is not possible to distinguish between the two energy injection progenitor scenarios, and as such energy injection can be parameterized such that $E_{\rm K, iso} \propto t^m$. In the 
magnetar central engine scenario $m = 1-q$, and in the stratified ejecta scenario $E_{\rm K, iso} \propto t^{3(s-1)/(7+s)}$, and $m = 3(s-1)/(7+s)$ \citep[for an ISM environment,][]{Zhang2006, lbm+2015ApJ...814....1L}. We find that a period of energy injection could explain the rapid rise observed at C-band due to the strong dependence of $F_{\nu, \rm m }$ on $m$. The flux evolution when $\nu_{\rm obs} < \nu_{\rm m}$ should follow $F_{\nu} \propto t^{(3+5m)/6}$ (see table 2 of \citealt{Zhang2006}). With the C-band rise following $\alpha \approx 2$, this would lead to $m \approx 1.8$. The subsequent rapid decay in the C-band light curve then requires a jet break at $t_{\rm jet}\approx19$~days.

In both the magnetar and stratified ejecta scenarios, the energy injection begins when the FS Lorentz factor slows down sufficiently for the slower material (characterized by $\Gamma_{\rm slow}$) to catch up and deposit its energy into the FS. In either case, we assume this collision is mild \citep{zhang2002}, such that $\Gamma_{\rm slow}$ is similar to the Lorentz factor of the FS ($\Gamma_{\rm FS}$), and therefore no RS is formed due to the collision. For our best-fit parameters, $\Gamma_{\rm FS}\approx4.2$ at the time of the start of the radio flare at $\approx11.2$~days and thus this scenario would imply a significant amount of energy is in material traveling at $\Gamma\lesssim4.2$.

To test this scenario, we construct a light curve incorporating energy injection within our modelling framework \citep[see:][]{lbm+2015ApJ...814....1L}. Using the best fit model parameters, we begin our energy injection period at $\delta t \approx 9~$days and end it at $\delta t \approx 19~$days, the time of the flare. We also set $t_{\rm jet} = 19.0~$days. During the injection period, we set the energy to be $E_{\rm K, iso} \propto t^{1.8}$, resulting in an increase by a factor of $\approx 3.8$. As a result of the earlier $t_{\rm jet}$ (compared to our initial FS fit), the energy injection model results in $\theta_{\rm jet} \approx 8.6^\circ$ and $E_{\rm K} \approx 3.4 \times 10^{51}~$erg, an factor of $\approx 1.3$ higher than the FS model alone (Section~\ref{sec:MCMC_fit_to_all}). We find this agrees well with the S-, C-, and X-band radio light curves (Figure~\ref{fig:FlarePlot}), however the L-band light curve is over-predicted by a factor of $\approx 2.9$ at $\delta t \approx 23.7~$days, and the X-ray light curve is similarly over-predicted by a factor of $\approx 2.2$ at $\delta t \approx 14.7~$days.

\subsection{Reverse Shock}
\label{sec:reverseshock}

We next consider an alternate scenario invoking an RS to explain the radio flare at $\approx 19~$days. An RS occurs either when: i) the ejecta interact with the surrounding environment or ii) when two ejecta shells collide \citep{zhang2002,l2017PhFl...29d7101Lyutikov, ld2018MNRAS.474.2813Lamberts}. In the former case, the RS dominates the afterglow immediately following the burst, whereas in the latter case there is a time delay before the RS dominates the afterglow. Given the timing of the radio flare at $\delta t \approx 19~$days in the afterglow of GRB\,210726A, the shell-collision model is the more relevant scenario. In this model, the X-ray re-brightening at $\delta t \approx 4.6~$days (see Section~\ref{sec:Xray_lc_temporal}) may be associated with the ejection of the second shell. Unlike the (mild) energy injection scenario (Section~\ref{sec:energyinjection}), this RS scenario requires a violent collision of the two shells \citep{zhang2002}. The conditions for a violent collision are that the second (inner, `injective') shell must be moving much faster than the first (outer, `impulsive') shell at the time of collision ($t_{\rm col}$), such that $\Gamma_{2} \gg \Gamma_{1}$. Additionally, we assume the energy of the second shell is $E_{2} < E_{1}$, and therefore the energy injected into the FS is not significant. Furthermore, any emission from the FS propagating in the outer shell from the collision of the two shells would be negligible compared to the RS emission \citep{zhang2002}.

An RS is formed in the inner shell at $t_{\rm col}$, and produces synchrotron radiation with its own set of break frequencies and peak flux: $\nu_{\rm sa, RS}$, $\nu_{\rm m, RS}$, $\nu_{\rm c, RS}$, $F_{\nu, \rm m, RS}$. The RS flux increases as the RS crosses the ejecta and decelerates the second shell at time, $t_{\rm dec}$. For a Newtonian RS, the evolution of the RS break frequencies after $t_{\rm dec}$ is set by the evolution of the Lorentz factor of the ejecta, such that $\Gamma \propto R^{-g}$, where $R$ is the radius of the ejecta shell \citep{mr1999MNRAS.306L..39M, ks2000ApJ...542..819Kobayashi}. In an ISM environment, numerical simulations suggest $g \approx 2.2$ \citep{ks2000ApJ...542..819Kobayashi}.

To test a scenario where the radio flare is caused by an RS produced in the collision of two ejecta shells, we construct a light curve incorporating an RS within our modeling framework (see: \citealt{lbz+2013ApJ...776..119L, lbm+2018ApJ...859..134Laskar}), assuming $g = 2.2$. We set $t_{\rm col} = 11.2~$days, the time of the first radio detection, and $t_{\rm dec} = 19~$days, the time of the radio flare. {In order to best match the radio light curves, we require $\nu_{\rm m, RS} < \nu_{\rm sa, RS}$. We find that an RS model with $F_{\nu, \rm m, RS} \approx 0.3~$mJy, $\nu_{\rm m, RS} \approx 6 \times 10^{8}~$Hz, $\nu_{\rm sa, RS} \approx 3.5 \times 10^{9}~$Hz, and $\nu_{\rm c, RS} \approx 2.4 \times 10^{14}~$Hz, provides a good match to the radio flare (Figure~\ref{fig:FlarePlot})}. This model over-predicts the L-band light curve by a factor of $\approx 1.7$; however, the fit to the higher-frequency radio observations is significantly improved compared to the FS model alone. {We caution that $t_{\rm col}$, $t_{\rm dec}$, $\nu_{\rm sa, RS}$, $\nu_{\rm m, RS}$, $\nu_{\rm c, RS}$, and  $F_{\nu, \rm m, RS}$ are not independently constrained due to our sparsely sampled data, and therefore additional combinations of these parameters may provide similarly good matches to the observations}. $\Gamma_2$ in an ISM environment follows \citep{lbm+2018ApJ...859..134Laskar}:

\begin{align}
    \Gamma_2 &= 2 \Gamma_1 \left(\frac{t_{\rm col}}{t_{\rm col}-\Delta t_{\rm L}}\right)^{1/2}
\end{align}
\noindent
where $\Gamma_1$ is the Lorentz factor of the first shell at $t_{\rm col}$ and $\Delta t_{\rm L}$ is the time of ejection of the second shell.
With $\Gamma_1 \approx 4.2$ at $t_{\rm col} = 11.2~$days and $\Delta t_{\rm L} = 4.6~$days, the time of the X-ray re-brightening, we find $\Gamma_2 \approx 10.9$. 

Like the energy injection scenario (Section~\ref{sec:energyinjection}), the RS model provides a reasonable match to the radio flare at $\approx19$~days and the subsequent $\gtrsim3$~GHz radio observations. The RS model also over-predicts the L-band light curve, although to a lesser degree. Unlike the energy injection model, the RS model does not over-predict the X-ray emission. While the RS model is not able to quite match the observed rapid rise at 6~GHz, we note that we use a fairly heuristic calculation of RS emission and that a more detailed calculation would require numerical simulations that are beyond the scope of this work. We consider the physical significance of both models in Sections~\ref{sec:energyinjectionimplications} and \ref{sec:RSimplications}.

\begin{figure}
    \centering
    \includegraphics[width = 0.9\columnwidth]{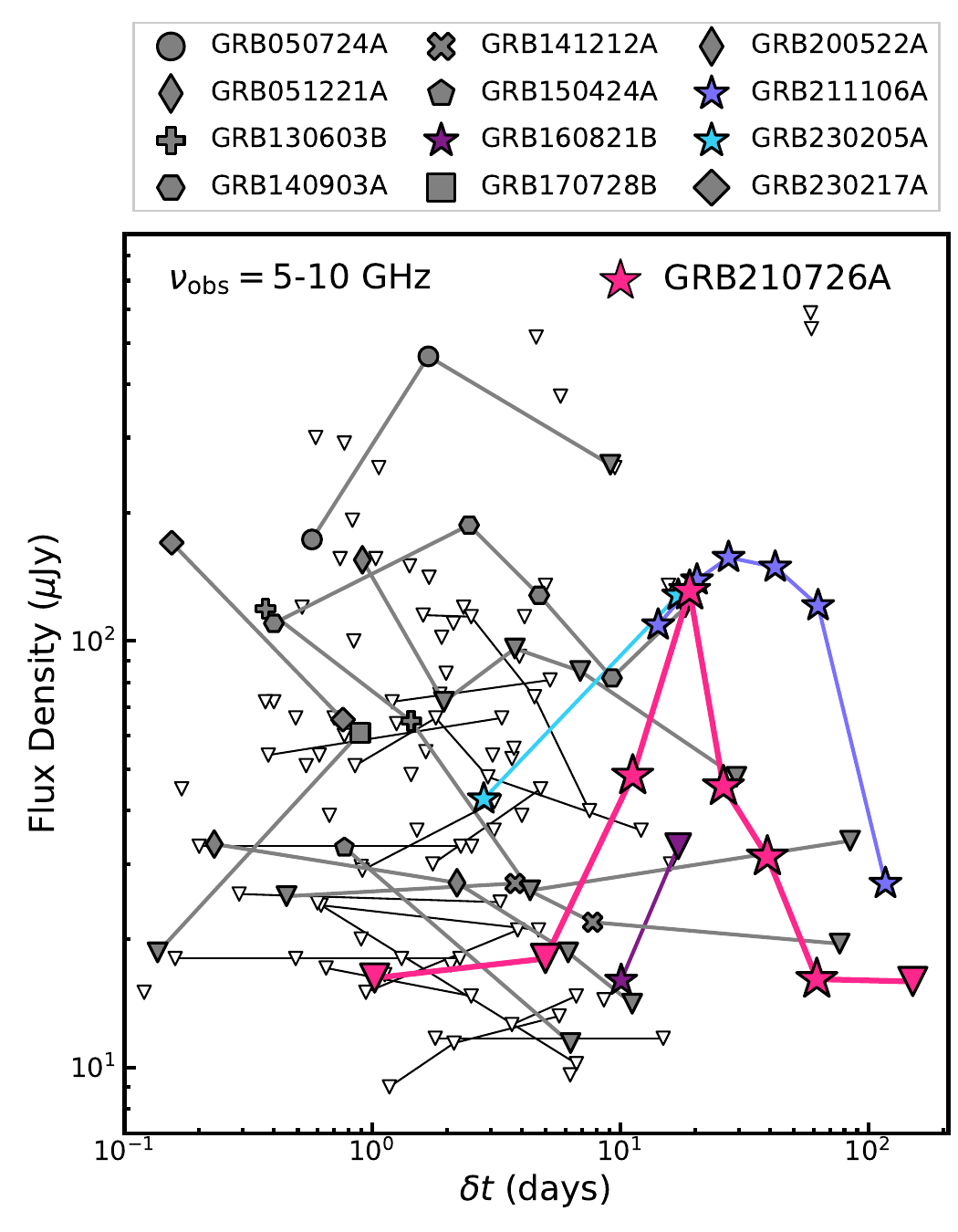}
    \caption{The light curves of all radio observations of short (or possibly short) GRBs at $\nu_{\rm obs} = 5$--$10~$GHz found in the literature. Triangles represent $3\sigma$ upper limits. Colored points indicate short GRBs that have radio detections at $\delta t \gtrsim 10~$days. Grey points indicate short GRBs with radio detections at $\delta t \lesssim 10~$days. Unfilled triangles represent short GRBs with no radio detections. }
    \label{fig:Radio_Observations}
\end{figure}

\section{Discussion}
\label{sec:Discussion}

\subsection{Implications of the Energy Injection and Reverse Shock Scenarios}
\label{sec:Implications}

\subsubsection{Energy Injection}
\label{sec:energyinjectionimplications}

Energy injection has been historically invoked to explain both long and short GRB afterglows that do not fit the standard FS afterglow model \citep[e.g.][]{2003Natur.426..138G, sbk+2006ApJ...650..261S, fx2006MNRAS.372L..19F, Ugarte2007, pmg+2009ApJ...696.1871P, rot+2010MNRAS.409..531R, hdm2012A&A...541A..88H, 
rom+2013MNRAS.430.1061R,
lbm+2015ApJ...814....1L, ltl+2019ApJ...883...48L, rhodes2021, dWLG2023arXiv230111985D_dewet, mmd+2023arXiv230600815M}. These studies include $> 20$ short GRBs for which some aspect of their afterglow was consistent with energy injection \citep{sbk+2006ApJ...650..261S, fx2006MNRAS.372L..19F, pmg+2009ApJ...696.1871P, rot+2010MNRAS.409..531R, hdm2012A&A...541A..88H, rom+2013MNRAS.430.1061R, ltl+2019ApJ...883...48L}, as well as some models of the multi-messenger binary NS merger, GRB\,170817A/GW\,170817  \citep{llt2020ApJ...899..105L, fkg+2022ApJ...933..243F}. 

In a study of 36 short GRBs with {\it Swift}-detected X-ray afterglows, \citet{rom+2013MNRAS.430.1061R} found that $\approx 50\%$ of the bursts were consistent with a magnetar central engine injecting energy into the X-ray afterglows. There, the majority of the excess energy was present at $\delta t \lesssim 1~$days, whereas we require energy injection at $\delta t > 1~$days for GRB\,210726A. Two previous short GRBs, GRB\,080503 and GRB\,160821B, displayed multi-wavelength re-brightenings at $\delta t \gtrsim 1~$days, with energy injection as the leading explanation \citep{pmg+2009ApJ...696.1871P, hdm2012A&A...541A..88H, ltl+2019ApJ...883...48L}. While GRB\,160821B had a radio afterglow re-brightening as well, the invocation of energy injection was constrained by only one radio detection, whereas GRB\,210726A exhibits a deviation from the FS model at multiple radio frequencies. 

\subsubsection{Reverse Shock}
\label{sec:RSimplications}

Radio emission from RSs has occasionally been found in afterglows of some long GRBs \citep{lbz+2013ApJ...776..119L, lab+2016, lbm+2018ApJ...859..134Laskar, lab+2018ApJ...862...94L, lvs+2019ApJ...884..121L, lag+2019ApJ...878L..26L, pcc+2014ApJ...781...37P, vdhpdb+2014MNRAS.444.3151V, alb+2017ApJ...848...69A, rvdhf+2020MNRAS.496.3326R, dtl+2022MNRAS.512.2337D, gg2023MNRAS.524L..78G, brf+2023NatAs.tmp..140B}. In almost all of these cases, the RS emission is assumed to peak soon after the GRB ($\delta t \approx 0~$s), however there are two GRBs that have exhibited multi-shell RSs: GRB\,030329 \citep{mmd+2023arXiv230600815M} and GRB\,140304A \citep{lbm+2018ApJ...859..134Laskar}, similar to the scenario invoked to explain the afterglow of GRB\,210726A.

To date, RSs have been invoked as a possible explanation for the multi-wavelength afterglow of four short (or possibly short) GRBs: 051221A, 160821B, 180418A, and 200522A \citep{sbk+2006ApJ...650..261S, ltl+2019ApJ...883...48L, tctb+2019MNRAS.489.2104T, bdw+2019ApJ...881...12B, flr+2021ApJ...906..127F, refv+2021ApJ...912...95R}. Notably, three of these short GRBs (GRB\,051221A, 160821B, 200522A) have radio afterglow detections which helped constrain the properties of the RS, similar to GRB\,210726A, whereas for GRB\,180418A the RS was only observed in the optical afterglow. However, all of the studies that have invoked RSs in short GRBs have required the RS to dominate at $\delta t \lesssim 1~$day, unlike GRB\,210726A, where the radio emission peaks at $\approx19$~days.

\subsubsection{Implications of a Long-Lived Central Engine}

Both the energy injection and RS scenarios invoke the presence of additional relativistic material interacting with the afterglow FS at $\gtrsim11.2$~days. In order to delay the time of the observed rapid radio brightening to several days, this material must be launched after the main ejecta that produces the GRB. In the stratified ejecta case for the energy injection model, this can be achieved by central engine activity past $T_{90}$, producing ejecta with Lorentz factors of $\Gamma\lesssim4.2$ (Section~\ref{sec:energyinjection}). This does not require extremely long central engine activity, as the timescale over which the light curves rise is set by the amount of time it takes the FS to decelerate to a Lorentz factor approaching that of the additional, stratified, unshocked ejecta, and not by the duration of engine activity. In the case of the RS, on the other hand, the production of a late-rising RS implicates a violent shell collision, which, in turn, requires engine activity at later times\footnote{Long-lasting RS propagating through dense ejecta shells have been proposed as a mechanism to produce X-ray flares, however the expected time scales are short ($\lesssim10^3$~s, \citealt{ld2018MNRAS.474.2813Lamberts}).}. If we associate the X-ray re-brightening at $\approx4.6~$days with the production of the latter shell, we would require the central engine to be active until at least that time {(a rest-frame time of $\approx 1.4$~days assuming $z = 2.38$)}. For a black hole central engine, late-time shell ejections could result from delayed (e.g.\ fallback) accretion \citep{rosswog2007MNRAS.376L..48R}, although the timescales appear extreme in this case.

If this central engine is a magnetar, one of the potential progenitors of the energy injection scenario, this scenario has possible testable predictions. If a magnetar remnant is produced from a NS-NS merger and remains stable against collapse, the spin down energy would be released into the surrounding medium, potentially producing synchrotron emission that peaks in the radio bands on the order of $\sim$months--years after the burst \citep{np2011Natur.478...82N, mb2014MNRAS.437.1821M, lgz2020ApJ...890..102L}. However, given the FS properties we derived in Section~\ref{sec:MCMC_fit_to_all}, in particular the circumburst density ($n_0 \approx 7.4\times 10^{-2}~{\rm cm}^{-3}$), it is not possible with current or planned radio facilities to place meaningful constraints on the existence of this central engine.

\subsection{Luminosity and Energetics}

We next explore the implications of the afterglow luminosity and energetics of GRB\,210726A. If the photometric redshift estimate of $z = 2.38$ is correct, GRB\,210726A is the highest known redshift short GRB with a detected radio afterglow \citep{BPC+2005Natur.438..988B, sbk+2006ApJ...650..261S, fbm+2014ApJ...780..118F, fbm+2015, 2017GCN.21395....1F, ltl+2019ApJ...883...48L, flr+2021ApJ...906..127F, lers+2022}, and among the highest redshift short GRBs known \citep{BRIGHT_II}. This is reflected in the radio luminosity of the burst, where the peak luminosity is $\approx 6.0 \times 10^{31}~$erg s$^{-1}$ Hz$^{-1}$, a factor of $\approx 5$ more luminous than the next most luminous event, GRB\,211106A (peak luminosity $\approx 1.1 \times 10^{31}~$erg s$^{-1}$ Hz$^{-1}$, assuming $z = 1.0$;  \citealt{lers+2022}). Additionally, GRB\,210726A is the most luminous short GRB in the X-rays at { a rest frame time of} $\delta t_{\rm rest} \gtrsim 0.1~$days ($L_{\rm (0.3-10)~keV} \approx 2.2 \times 10^{46}~{\rm erg~s^{-1}}$ at $\delta t_{\rm rest} \approx 0.1~$days). The high redshift also is reflected in the large energetics of the GRB. Assuming only the FS solution (e.g. no energy injection component), GRB\,210726A is among the most energetic short GRBs to date. Two additional GRBs\,180418A and 211106A have similar or higher inferred beaming-corrected kinetic energies than GRB\,210726A, with $E_{\rm K} \approx 2$--$6 \times 10^{51}$~erg \citep{lers+2022,refv+2021ApJ...912...95R}; however both of these bursts had even more limited host galaxy data making their redshifts less secure than GRB\,210726A \citep{BRIGHT_II}. 

At $z = 2.38$, the isotropic $\gamma$-ray energy of GRB\,210726A is $E_{\gamma, \rm iso} \approx 5.9 \times 10^{50}~$erg, compared to the much higher $E_{K, \rm iso}  \approx 8.1 \times 10^{52}~$erg. We calculate the radiative efficiency $\eta_\gamma = E_{\gamma, \rm iso} / (E_{\gamma, \rm iso} + E_{K, \rm iso})$ and find $\eta_\gamma \approx 7.2 \times 10^{-3}$. Assuming the energy injection scenario (Section~\ref{sec:energyinjection}) is correct, this is consistent with the findings of \citet{lbm+2015ApJ...814....1L}, which showed that GRBs with evidence of energy injection display low radiative efficiencies, due to the prompt $\gamma$-ray emission being dominated by only the fastest moving ejecta. 

We also explore the implications if GRB\,210726A instead originated closer to the median redshift of short GRBs \citep[$ z = 0.64$, ][]{BRIGHT_I, BRIGHT_II}, given that the redshift of $z \approx 2.4$ is only based on photometric data. In this case, GRB\,210726A would be the second most luminous in the radio, with a peak luminosity of $\approx 2.4\times 10^{30}~$erg s$^{-1}$ Hz$^{-1}$, though its X-ray luminosity would be unremarkable for a short GRB ($L_{\rm (0.3-10)~keV} \approx 8.7 \times 10^{44}~{\rm erg~s^{-1}}$ at $\delta t_{\rm rest} \approx 0.1~$days). We further find that $E_{\rm K} \approx 5.2 \times 10^{50}~$erg at $z = 0.64$, still placing GRB\,210726A among the most energetic short GRBs with measured beaming-corrected kinetic energies.

\subsection{The Importance of Long-term Radio Monitoring}

The radio afterglow of GRB\,210726A was detected relatively late ($\delta t \approx 11.2~$days), despite several epochs of observations prior, all of which yielded non-detections. It is clear that the radio afterglow detection, let alone the discovery of the radio flare, was only possible with continued monitoring at $\delta t \gtrsim 10~$days. Thus, we now explore the importance of extending radio observational campaigns to late times for short GRBs. We gather all of the published radio observations of short GRBs\footnote{We include possibly short GRBs, such as GRB\,230205A \citep{2023GCN.33372....1S_230205A}, as well as long GRBs from compact object mergers, such as GRB\,211211A \citep{rgl+2022Natur.612..223R}. We exclude GRB\,170817A as its orientation was off-axis and thus not directly comparable.} (\citealt{BPC+2005Natur.438..988B, sbk+2006ApJ...650..261S, fbm+2014ApJ...780..118F, fbm+2015} and references therein, \citealt{2017GCN.21395....1F, ltl+2019ApJ...883...48L, tctb+2019MNRAS.489.2104T, flr+2021ApJ...906..127F, lers+2022, 2023GCN.33309....1S_230205A, 2023GCN.33372....1S_230205A, 2023GCN.33358....1S_230217A, 230217A_ATCA_2023GCN.33433....1A}), and plot the C- and X-band observations in Figure~\ref{fig:Radio_Observations}. It is clear that there is a distinct lack of radio observations at $\delta t \gtrsim 10~$days for nearly all short GRBs, with only 15 of the 86 in our sample having radio observations at $\delta t \gtrsim 10~$days in any radio frequency, representing missed opportunities. Indeed, the detection efficiency is somewhat higher for those with such late observations: $4/15$ ($\approx 27\%$) have a detected radio afterglow at $\delta t \gtrsim 10~$days, with 3 of the 4 being detected in the last $\sim 2~$years (GRBs\,210726A, 211106A, 230205A), demonstrating we have only recently recognized the importance of extended radio campaigns for short GRBs.

The radio afterglows of short GRBs are generally expected to fall below the peak frequency ($\nu_{\rm radio} < \nu_{\rm m}$), and thus, the  light curve should rise as $F_\nu \propto t^{1/2}$ until $\nu_{\rm radio} = \nu_{\rm m}$ or a jet break occurs \citep{Rhoads1999, SariPiranHalpern1999, GS2002}. Therefore, it is not unexpected to have a radio light curve that rises and becomes detectable at later times (e.g., $\delta t \gtrsim 10~$days\footnote{{While the observed rise time of a radio afterglow is dependent on the redshift of the GRB, the redshift of a short GRB is often unknown at the time of detection, given the faintness of the optical afterglows and hosts \citep{2022MNRAS.515.4890O, BRIGHT_I, BRIGHT_II}. We therefore use $\delta t \approx 10~$days as a general benchmark, which encompasses a rest frame time of $\approx 3$--$9$~days for the typical short GRB population.}}). Detecting the radio afterglow helps constrain the location of $\nu_{\rm sa}$ and $\nu_{\rm m}$, effectively constraining $E_{\rm K,iso}$ and $n_{0}$ and therefore breaking degeneracies that can occur within these parameters from only constraining the location of $\nu_{\rm c}$ \citep{fbm+2015}. With only 13 short GRBs with radio afterglows to date, each additional detection is integral to our understanding of the energetics and environments in which these bursts occur. Therefore, we conclude that adjusting radio observation strategies to cover a longer expected rise time of a radio afterglow will benefit the holistic understanding of short GRBs.

\section{Conclusions}
\label{sec:Conclusions}

We have presented multiwavelength observations of short GRB\,210726A. We demonstrate that an FS alone is not sufficient to explain all of the features of the afterglow and in particular the radio flare. We  presented additional emission mechanisms to explain the radio flare of GRB\,210726A. We have come to the following conclusions:

\begin{enumerate}
    \item At $z \sim 2.4$, GRB\,210726A is among the most distant short GRBs with a determined redshift to date, and the highest redshift short GRB with a detected radio afterglow. Consequently, GRB\,210726A is the most luminous short GRB in both the radio and the X-rays.
    \item The 6~GHz afterglow displays a rapid rise from $\approx 11$--$19~$days after the burst. Additionally, the $3$--$10$~GHz afterglow display a rapid decline $\approx 19~$days after the burst. This radio flare is best explained by an energy injection event or an RS. 
    \item If the radio flare is from energy injection, the resulting energy injection event results in a a higher $E_{\rm K, iso}$ by a factor of $\approx 3.8$ and a higher $E_{\rm K}$ by a factor of $\approx 1.3$, compared to the FS model. If instead the radio flare is from an RS, the RS would have been produced by the collision of two ejecta shells, where we assume the X-ray re-brightening at $\delta t \approx 4.6~$days is attributed to the ejection of the second shell. The Lorentz factor of the second shell is a factor of $\approx 2.6$ higher than the initial shell.
    \item While both the energy injection scenario and the RS scenario match the observed $\gtrsim 3~$GHz behavior at $\gtrsim 19~$days, both scenarios over-predict the $1.3~$GHz afterglow. Additionally, the energy injection scenario over-predicts the X-ray afterglow, whereas the RS scenario may miss the rise of the 6~GHz afterglow. However, the RS scenario may require extreme late time central engine activity to produce the second shell. Therefore, it is unclear which scenario is more likely to have produced the radio flare.
\end{enumerate}

Our work demonstrates the importance of continued radio monitoring of short GRBs, especially after $\delta t \gtrsim 10~$days, as the radio afterglow of GRB\,210726A would have been missed if radio monitoring has ceased by $\approx 10~$days. This is especially important given the paucity of short GRB radio afterglow detections, which are crucial for breaking degeneracies in afterglow models to better understand the energetics and environments of these neutron star merger progenitors. Additionally, given that GRB\,210726A is the first short GRB detected with MeerKAT, our work highlights the strength of multi-frequency radio observations, especially at low frequencies. 

\section{Acknowledgements}

G.S. acknowledges support for this work was provided by the NSF through a Student Observing Support award from the NRAO. The Fong Group at Northwestern acknowledges the support of the National Science Foundation under grant No. AST-1909358 and CAREER grant No. AST2047919. Support for this work was in part provided by the National Aeronautics and Space Administration through Chandra Award Numbers \#DD1-22131X and \#GO1-22043X issued by the Chandra X-ray Center, which is operated by the Smithsonian Astrophysical Observatory for and on behalf of the National Aeronautics Space Administration under contract NAS8-03060. W.F. gratefully acknowledges the support of the David and Lucile Packard Foundation, the Alfred P. Sloan Foundation, and the Research Corporation for Science
Advancement through Cottrell Scholar Award \#28284. C.D.K. acknowledges partial support from a CIERA postdoctoral fellowship. A.R.E. is supported by the European Space Agency (ESA) Research Fellowship. E.B. acknowledges support from NSF and NASA grants. This study was enabled by a Radboud Excellence fellowship from Radboud University in Nijmegen, Netherlands. I.H. acknowledges support of the Science and Technology Facilities Council (STFC) grants [ST/S000488/1] and [ST/W000903/1]. I.H. acknowledges support from a UKRI Frontiers Research Grant [EP/X026639/1], which was selected by the European Research Council. I.H. acknowledges support from the South African Radio Astronomy Observatory which is a facility of the National Research Foundation (NRF), an agency of the Department of Science and Innovation. I.H. acknowledges support from Breakthrough Listen. Breakthrough Listen is managed by the Breakthrough Initiatives, sponsored by the Breakthrough Prize Foundation. D.B.M. is supported by the European Research Council (ERC) under the European Union’s Horizon 2020 research and innovation programme (grant agreement No.~725246). The Cosmic Dawn Center is supported by the Danish National Research Foundation.
P.V. acknowledges funding
from NASA cooperative agreement 80MSFC22M0004. N.R.T. acknowledges support from STFC grant ST/W000857/1.

This work made use of data supplied by the UK Swift Science Data Centre at the University of Leicester. The National Radio Astronomy Observatory is a facility of the National Science Foundation operated under cooperative agreement by Associated Universities, Inc. VLA observations for this study were obtained via projects VLA/20B-057 and VLA/21B-198. This work is partly based on observations collected at the European Southern Observatory under ESO programme 110.24CF.003. We also acknowledge the staff who operate and run the AMI-LA telescope at Lord’s Bridge, Cambridge, for the AMI-LA radio data. AMI is supported by the Universities of Cambridge and Oxford, and acknowledges support from the European Research Council under grant ERC-2012-StG-307215 LODESTONE. e-MERLIN, and formerly, MERLIN, is a National Facility operated by the University of Manchester at Jodrell Bank Observatory on behalf of the STFC. We acknowledge Jodrell Bank Centre for Astrophysics, which is funded by the STFC. The MeerKAT telescope is operated by the South African Radio Astronomy Observatory, which is a facility of the National Research Foundation, an agency of the Department of Science and Innovation.

This research was supported in part through the computational resources and staff contributions provided for the Quest high performance computing facility at Northwestern University which is jointly supported by the Office of the Provost, the Office for Research, and Northwestern University Information Technology. W. M. Keck Observatory and MMT Observatory access was supported by Northwestern University and the Center for Interdisciplinary Exploration and Research in Astrophysics (CIERA). Some of the data presented herein were obtained at the W. M. Keck Observatory, which is operated as a scientific partnership among the California Institute of Technology, the University of California and the National Aeronautics and Space Administration. The Observatory was made possible by the generous financial support of the W. M. Keck Foundation. The authors wish to recognize and acknowledge the very significant cultural role and reverence that the summit of Maunakea has always had within the indigenous Hawaiian community. We are most fortunate to have the opportunity to conduct observations from this mountain. Based on observations obtained at the international Gemini Observatory (Program IDs GS-2021A-Q-112, GS-2023A-FT-101), a program of NOIRLab, which is managed by the Association of Universities for Research in Astronomy (AURA) under a cooperative agreement with the National Science Foundation on behalf of the Gemini Observatory partnership: the National Science Foundation (United States), National Research Council (Canada), Agencia Nacional de Investigaci\'{o}n y Desarrollo (Chile), Ministerio de Ciencia, Tecnolog\'{i}a e Innovaci\'{o}n (Argentina), Minist\'{e}rio da Ci\^{e}ncia, Tecnologia, Inova\c{c}\~{o}es e Comunica\c{c}\~{o}es (Brazil), and Korea Astronomy and Space Science Institute (Republic of Korea).

\facilities{VLA, MeerKAT, \textit{e}-Merlin, AMI-LA, Swift, Fermi, Chandra, Keck:LRIS, Keck:MOSFIRE, MMT:MMIRS, Gemini-North:GMOS, Gemini-South:GMOS, VLT:MUSE}
\software{CASA \citep{CASA}, pwkit \citep{Pwkit}, scipy \citep{2020SciPy-NMeth}, matplotlib \citep{matplotlib}, pandas \citep{pandas}, numpy \citep{numpy}, Prospector \citep{Leja_2017}, PyPeIt \citep{phw+2020}, IRAF \citep{iraf1, iraf2}, Xspec \citep{Xspec_1996ASPC..101...17A}, CIAO \citep[v.4.12][]{Fruscione2006}, emcee \citep{emcee_2013PASP..125..306F}}

\clearpage
\newpage
\appendix
\restartappendixnumbering

\section{Extraction of X-ray Afterglow and Contaminants. Spectral Analysis Configuration}
\label{sec:Xray_contaminants}

We extract the spectrum of the afterglow from a $2\arcsec$-radius circular region centered at the CXO afterglow position, and the background from a source-free $7\arcsec-15\arcsec$ radii annulus. For the contaminants, we use the same size source and background regions, except for X2 and X3 with background annuli of $9\arcsec$--$18\arcsec$ radii. We generate all sources and background spectra, and ancillary response and redistribution matrix files with the \texttt{specextract} tool. We use {\tt Xspec} (12.10.1f; \citealt{Xspec_1996ASPC..101...17A}) to fit all the spectra and calculate the X-ray unabsorbed fluxes ($0.3$--$10$\,keV). We bin the spectra to ensure one count per bin and, choose \texttt{WILM} (\citealt{wilms2000}) abundances, \texttt{VERN} (\citealt{Verner1996}) X-ray cross-sections and W-statistics for background-subtracted Poisson data (\citealt{Wachter1979}).

We model the CXO spectra of X1, X2, and X3 using a single absorbed power-law model (\texttt{tbabs x pow}). We find no evidence for spectral evolution of the three contaminating sources and present the best-fit spectral parameters tied across the full CXO dataset for these three sources in Table~\ref{tab:data_xray_contaminants}. 

\begin{deluxetable}{cccc}[t!]
 \centering
 \tabletypesize{\footnotesize}
 \tablecolumns{4}
 \tablecaption{X-ray spectral parameters of contaminants}
 \tablehead{
    \colhead{Source Name} &
   \colhead{$N_H$} &
   \colhead{$\Gamma_{X}$} &
   \colhead{Unabsorbed Flux}\\[-0.1in]
   \colhead{(cm$^{-2}$)} &
   \colhead{} &
   \colhead{($10^{-14}$\,erg\,s$^{-1}$\,cm$^{-2}$)}
   }
\startdata 
X1 & $0.03_{-0.03}^{+1442.04} \times 10 ^{18}$ & $1.2_{-0.1}^{+0.2}$ & $3.8\pm0.3$   \\
X2 & $0.01_{-0.01}^{+114.63} \times 10 ^{20}$ & $1.5_{-0.6}^{+1.1}$ & $1.2\pm0.2$   \\
X3 & $0.6_{-0.6}^{+1.5} \times 10 ^{22}$ & $2.3_{-0.7}^{+0.9}$ & $1.5_{-0.2}^{+0.3}$   \\
\enddata
\tablecomments{Time is log-centered (observer frame). Fluxes are reported in the 0.3--10\,keV band. Uncertainties are $1\sigma$}
\label{tab:data_xray_contaminants}
\end{deluxetable}

\section{Forward Shock Fit while Masking the MeerKAT Observations}
\label{sec:MeerKAT_masked}

\begin{deluxetable}{c|cc}
 \tabletypesize{\footnotesize}
 \tablecolumns{3}
 \tablecaption{Forward Shock Parameters, masking the $1.3~$GHz Afterglow}
 \tablehead{   
   \colhead{Parameter} &
   \colhead{Best Fit Model} &
   \colhead{MCMC Results} 
   }
 \startdata 
$p$ & $2.03$ & $2.05^{+0.03}_{-0.02}$ \\
$E_{\rm K, iso}~(\times 10^{52}~{\rm erg})$ & $5.6$ & $4.1^{+2.8}_{-2.1}$ \\
$n_0~({\rm cm}^{-3})$ & $8.4\times 10^{-2}$ &$6.5^{+7.2}_{-3.7} \times 10^{-2}$ \\
$\epsilon_{\rm e}$ & $9.8\times 10^{-1}$ &$5.9^{+2.5}_{-2.1} \times 10^{-1}$ \\
$\epsilon_{\rm B}$ & $4.0\times 10^{-4}$ &$5.1^{+32.7}_{-3.6} \times 10^{-4}$ \\
$t_{\rm jet}~({\rm days})$ & $47.3$ & $58.9^{+42.9}_{-22.6}$ \\
$\theta_{\rm jet}~({\rm deg})$ & $15.2$ & $16.6^{+3.7}_{-2.8}$ \\
$E_{\rm K}~(\times 10^{52}~{\rm erg})$ & $1.9\times 10^{-1}$ &$1.7^{+1.4}_{-0.9} \times 10^{-1}$ \\
$A_{V}~(\rm mag)$ & $\gtrsim 0.88$ & --\\
\enddata
\tablecomments{\emph{Top:} The best-fit (left) and summary statistics (median and 68\% credible intervals, right) parameters from the marginalized posterior density functions of the forward shock (FS) afterglow parameters from our MCMC modeling, excluding the MeerKAT $1.3~$GHz afteglow. The parameters of the best-fit model may differ from the summary statistics as the former is the peak of the likelihood distribution and the latter is calculated from the full marginalized posterior density functions of each parameter.}
\label{tab:FS_fit_noMK}
\end{deluxetable}

\begin{figure}
    \centering
    \includegraphics[width = 0.48\textwidth]{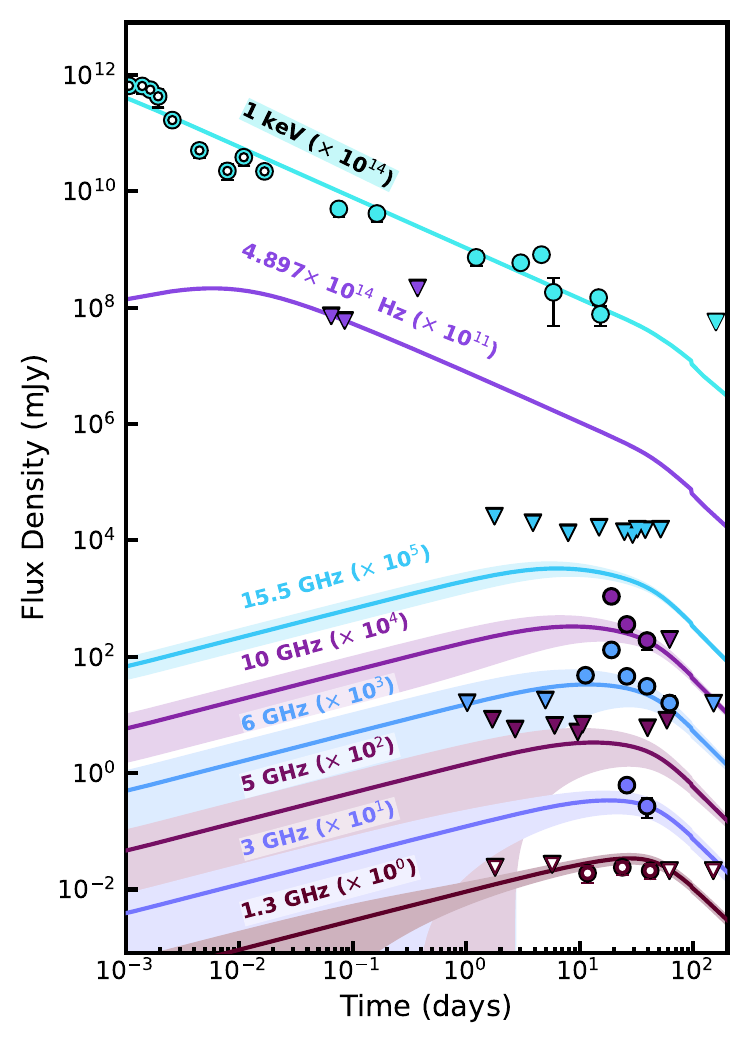}
    \caption{Forward shock (FS) fit of the afterglow of GRB\,210726A, with the MeerKAT $1.3~$GHz afterglow masked. Circles indicate detections, whereas triangles indicate $3 \sigma$ limits. Open circles indicate data not included in the fit. Shaded regions indicate predicted variability due to scintillation.}
    \label{fig:forwardshockfits_noMK}
\end{figure}

We fit the the afterglow of GRB\,210726A with the same procedure mapped out in Section~\ref{sec:MCMC_fit_to_all}, but with the L-band observations masked. The resultant fit produced a slightly lower (better) $\chi^2/d.o.f. \approx 64/24 \approx 2.7$, but also lower (worse) $\mathcal{L} \approx 103$. Overall, the parameters do not change significantly from the fit with the full data set, other than $\epsilon_{\rm B}$, which increased by a factor of $\approx 3.6$ (Table~\ref{tab:FS_fit_noMK}). We find that the fit is only marginally improved for the high-frequency data, with deviations by factors of $\approx 1.2$--$1.6$ between the C-band model and data ($\approx 4.2$ for the radio flare; Figure~\ref{fig:forwardshockfits_noMK}). However, the L-band observations are overpredicted by a factor of $\approx 1.5$.

One possible mechanism for suppressing the lower frequency emission is additional synchrotron self-absorption opacity from non-accelerated (``thermal'') electrons behind the FS ($f_{\rm NT}<1$).  Such a scenario can lead to an increase in $\nu_{\rm sa}$ by a factor of $\approx 10-100$ \citep{ressler_laskar2017ApJ...845..150R, wbi2018MNRAS.480.4060W}. A higher value of $\nu_{\rm sa}\gtrsim1.4$~GHz would decrease the L-band model flux. Similar arguments have been made previously \citep[see e.g.][]{dWLG2023arXiv230111985D_dewet}. This is supported by some simulations, which have shown $f_{\rm NT}$ to be as low as $\approx 0.01$--$0.15$ \citep{ss2011}. Additionally, this scenario can also help alleviate the high value of $\epsilon_{\rm e}$ found for this burst. A study of the radio peak of GRB afterglows indicated that $0.01 < \epsilon_{\rm e} < 0.2$ and $0.1 < f_{\rm NT} < 1$, with the two parameters being positively correlated
\citep{2023MNRAS.518.1522D_DvdHB2023}. Therefore, the high values of $\epsilon_{\rm e} \approx 0.90$--$0.98$ found in our MCMC afterglow fits is consistent with a situation where $f_{\rm NT} < 1$. We note that all the parameters in our modeling framework are degenerate with $f_{\rm NT}$ \citep{ew2005} and that similar arguments have been made previously (e.g. \citealt{lab+2016, lbc+2018, kf2021, lers+2022, slf+2022}). 

{A consequence of this degeneracy is that if $f_{\rm NT} < 1$, this implies the true value of $\epsilon_{\rm B}$ ($\approx 4 \times 10^{-4}$ assuming $f_{\rm NT} = 1$) would be reduced by the same factor. Such a low value of $\epsilon_{\rm B}$ is not a concern, as several studies of GRB afterglows have found wide distributions for $\epsilon_{\rm B}$, with median values as low as $\sim 10^{-5}$ \citep[e.g.][]{2013MNRAS.435.3009L, 2014ApJ...785...29S, 2014MNRAS.442.3147B}. Additionally, $E_{\rm K, iso}$ (and therefore $E_{\rm K}$) and $n_0$ are inversely affected by $f_{\rm NT}$, and the true value would be higher than the values we derive in our fits. The increase in $n_0$ is not concerning, as our derived value is still well within the typical range for short GRBs \citep{fbm+2015}. However, the derived value of $E_{\rm K}$ is already quite high, and any increase would further support the conclusion that GRB\,210726A one of the most energetic short GRBs detected to date.}

\section{Alternative Explanations for the Radio Flare}
\label{sec:extrinsicscenarios}

Here we explore three alternative explanations for the radio flare of short GRB\,210726A: scintillation, an off axis afterglow, and a variable external medium.

\subsection{Scintillation}\label{sec:DISS}

\begin{deluxetable}{ccccc}
 \tabletypesize{\footnotesize}
 \tablecolumns{5}
 \tablecaption{VLA Spectral Variability Investigation}
 \tablehead{   
   \colhead{$\delta t^{\rm{a}}$} &
   \colhead{$\nu^{\rm{b}}$} &
   \colhead{Bandwidth$^{\rm{c}}$} &
   \colhead{Flux density} &
   \colhead{RMS}\\
   \colhead{(days)} &
   \colhead{(GHz)} &
   \colhead{(MHz)} &
   \colhead{($\mu$Jy)} &
   \colhead{($\mu$Jy)}
   }
 \startdata 
11.2	& 4.74 	& 512 	& 31$\pm$12 	& 8 \\
	& 5.26 	& 512 	& 50$\pm$15 	& 10 \\
	& 6.84 	& 512 	& 56$\pm$9 	& 6 \\
	& 7.36 	& 512 	& 70$\pm$12 	& 8 \\
\hline
	& 4.99 	& 1024 	& 41$\pm$9 	& 6 \\
	& 7.10 	& 1024 	& 63$\pm$8 	& 5 \\
\hline
19.0	& 4.62 	& 256 	& 98$\pm$18 	& 12 \\
	& 4.87 	& 256 	& 106$\pm$22 	& 15 \\
	& 5.13 	& 256 	& 114$\pm$14 	& 10 \\
	& 5.38 	& 256 	& 101$\pm$21 	& 15 \\
	& 6.72 	& 256 	& 159$\pm$16 	& 12 \\
	& 6.97 	& 256 	& 150$\pm$13 	& 9 \\
	& 7.23 	& 256 	& 170$\pm$18 	& 13 \\
	& 7.48 	& 256 	& 149$\pm$17 	& 12 \\
\hline
	& 4.99 	& 1024 	& 103$\pm$10 	& 7 \\
	& 7.10 	& 1024 	& 155$\pm$8 	& 5 \\
\hline
\hline
19.0	& 8.17 	& 256 	& 122$\pm$15 	& 10 \\
	& 8.42 	& 256 	& 104$\pm$21 	& 15 \\
	& 8.68 	& 256 	& 121$\pm$12 	& 9 \\
	& 8.93 	& 256 	& 124$\pm$16 	& 11 \\
	& 10.62 	& 256 	& 71$\pm$18 	& 13 \\
	& 10.87 	& 256 	& 118$\pm$23 	& 16 \\
	& 11.27 	& 256 	& 104$\pm$16 	& 11 \\
	& 11.38 	& 256 	& 123$\pm$18 	& 12 \\
\hline
	& 8.55 	& 1024 	& 119$\pm$8 	& 6 \\
	& 11.0 	& 1024 	& 99$\pm$9 	& 7 \\
\hline
\hline
26.0	& 2.5 	& 1024 	& 66$\pm$36 	& 25 \\
	& 3.5 	& 1024 	& 70$\pm$11 	& 8 \\
	& 5.0 	& 1024 	& 44$\pm$14 	& 10 \\
	& 7.1 	& 1024 	& 45$\pm$10 	& 7 \\
	& 8.55 	& 1024 	& 27$\pm$11 	& 8 \\
	& 11.0 	& 1024 	& 49$\pm$11 	& 8 \\
\enddata
\tablecomments{$^{\rm{a}}$ Mid-time of entire VLA observation since {\it Swift}/XRT trigger. $^{\rm{b}}$ Central frequency. $^{\rm{c}}$ Width of frequency bin.}
\label{tab:VLA_epoch3_4_5_frequency}
\end{deluxetable}

\begin{deluxetable}{cccccc}
 \tabletypesize{\footnotesize}
 \tablecolumns{6}
 \tablecaption{VLA Temporal Variability Investigation}
 \tablehead{   
   \colhead{Date} &
   \colhead{Mid-time$^{\rm{a}}$} &
   \colhead{Bin Width$^{\rm{b}}$} &
   \colhead{$\nu^{\rm{c}}$} &
   \colhead{Flux density} &
   \colhead{RMS}\\
   \colhead{} &
   \colhead{ (UTC)} &
   \colhead{(s)} &
   \colhead{(GHz)} &
   \colhead{($\mu$Jy)} &
   \colhead{($\mu$Jy)}
   }
 \startdata 
14-Aug-2021	& 19:30:55.0 	& 510 	& 6.0 	& $125 \pm 15$ & $11$ \\
		& 19:40:57.5 	& 510 	& 	& $131 \pm 19$ & $14$ \\
		& 19:51:00.0 	& 510 	& 	& $113 \pm 19$ & $14$ \\
		& 20:01:05.0 	& 510 	& 	& $139 \pm 20$ & $14$ \\
		& 20:11:07.5 	& 510 	& 	& $120 \pm 22$ & $16$ \\
		& 20:21:10.0 	& 510 	& 	& $122 \pm 17$ & $12$ \\
		& 20:31:15.0 	& 510 	& 	& $136 \pm 15$ & $11$ \\
		& 20:41:17.5 	& 510 	& 	& $132 \pm 17$ & $12$ \\
\hline
14-Aug-2021	& 18:11:31.5 	& 510 	& 9.8 	& $90 \pm 17$ & $12$ \\
		& 18:21:15.0 	& 510 	& 	& $102 \pm 16$ & $11$ \\
		& 18:30:57.0 	& 510 	& 	& $109 \pm 15$ & $11$ \\
		& 18:40:40.5 	& 510 	& 	& $96 \pm 16$ & $11$ \\
		& 18:50:24.0 	& 510 	& 	& $120 \pm 17$ & $12$ \\
		& 19:00:07.5 	& 510 	& 	& $121 \pm 14$ & $10$ \\
		& 19:09:51.0 	& 510 	& 	& $104 \pm 16$ & $12$ \\
		& 19:19:34.5 	& 510 	& 	& $112 \pm 12$ & $9$ \\
  \enddata
\tablecomments{$^{\rm{a}}$ Mid-time of the time bin. $^{\rm{b}}$ Width of time bin. $^{\rm{c}}$ Central Frequency.}
\label{tab:VLA_epoch4_time}
\end{deluxetable}

First, we investigate whether interstellar scintillation could be responsible for the apparent late and fast rising C-band afterglow. Evidence of scintillation has been discovered in GRB afterglows before, leading to extreme temporal and spectral variability (e.g. \citealt{gbw+2018A&A...614A..29G, alb+2019, 2022MNRAS.513.1895R, 2022arXiv221111212A}). 
Specifically, refractive and diffractive insterstellar scintillation (RISS and DISS, respectively), have the strongest effects at early times when the angular size of the GRB jet is small \citep{1997NewA....2..449G, 1998MNRAS.294..307W, 1997NewA....2..449G, granot_vanderhorst_2014PASA...31....8G}. While RISS and DISS can individually produce maximum flux modulations of a factor of 2, our afterglow model under predicts the radio flare by a factor of $\sim 4$--$5$. However, we still investigate here to see if it can reconcile some of the discrepancy between our afterglow model and our observations. 

According to the NE2001 model \citep{NE2001}, observations below a transition frequency of 6.97\,GHz are in the strong scattering regime for the coordinates of GRB\,210726A, within the band of our C-band observations. At the lowest edge of the band, we would expect to see temporal and/or spectral variations of flux on the timescales of 1.2 (5.2) hours, with flux changes on the order of 2.0 (1.8) for DISS (RISS).

\begin{figure}
    \centering
    \includegraphics[width = 0.45\textwidth]
    {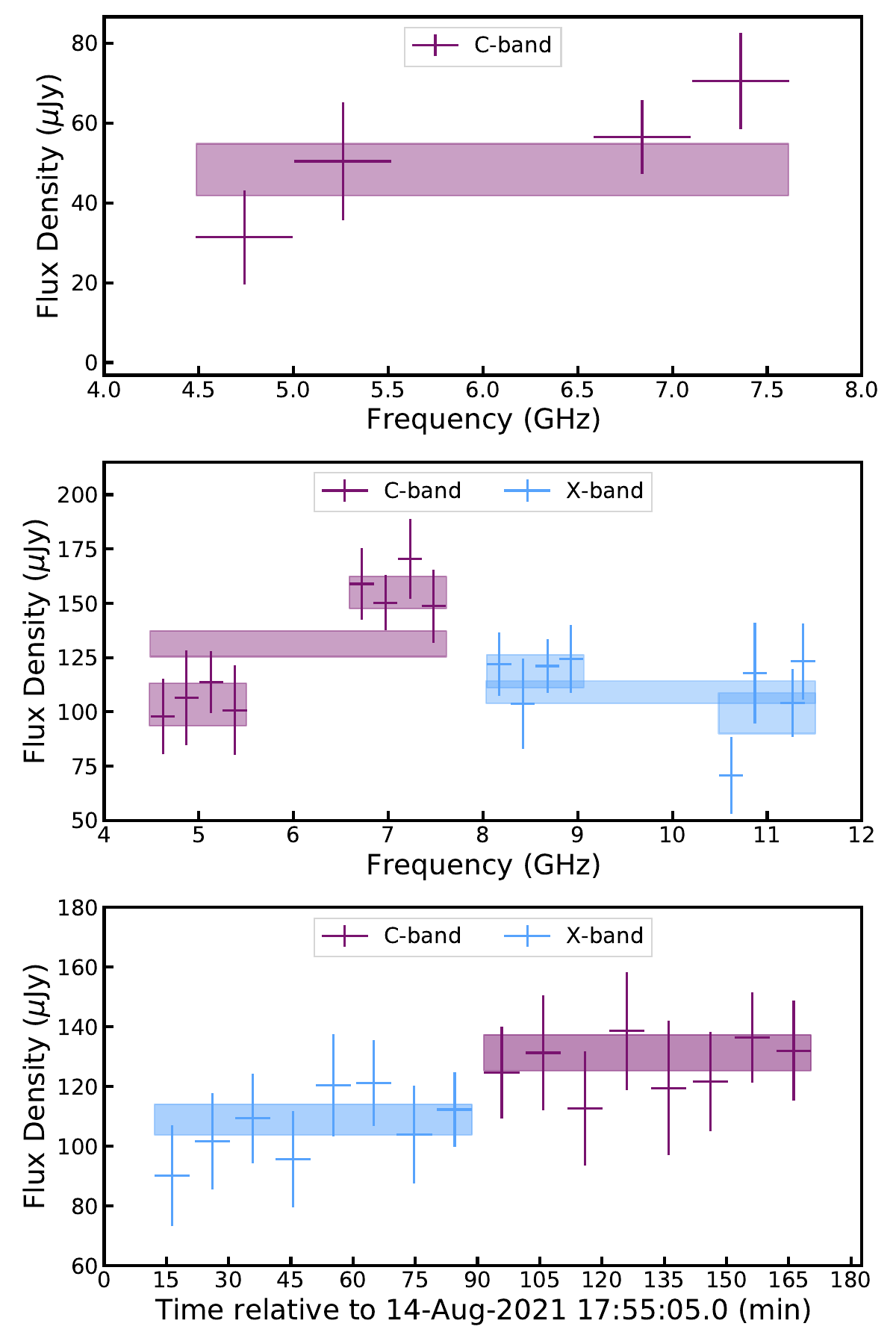}
    \caption{{\it Top}: The 11.2 day VLA C-band observation, broken up into four 256 MHz bins (error bars indicate the 1$\sigma$ flux density and width of the frequency bin). The shaded region indicates the flux density of the full C-band bandwidth at 1$\sigma$, with the width of the shaded region indicating the bandwidth. 
    {\it Middle}: The 19.0 day VLA C-band (purple points) and X-band (blue points) observation, broken up into sixteen 128 MHz bins (eight bins per band, error bars indicate the 1$\sigma$ flux density and width of the frequency bin). The shaded regions indicates the flux density of the full C-band (purple) and X-band (blue) bandwidth,  at 1$\sigma$, as well as the flux density of the lower and upper side bands, the width of the shaded region indicating the bandwidth. {\it Bottom}: The 19.0 day VLA C-band (purple points) and X-band (blue points) observation, broken up into sixteen $\sim 510~$s time bins (eight bins per band, error bars indicate the 1$\sigma$ flux density and width of the time bin). The shaded region indicates the flux density of the full C-band (purple) and X-band (blue) bandwidth at 1$\sigma$, with the width of the shaded region indicating the temporal coverage of the observation at that band.}
    \label{fig:DISS}
\end{figure}

We search for evidence of DISS in the both C- and X-band observations, as our afterglow models are most discrepant with these bands at the time of the radio flare. Each of our C-band and X-band observations with the VLA last approximately 1.5\,hours, comparable to the timescale of DISS, which is a narrowband phenomena that can lead to spectral variability. We split the earliest C-band detection at $\delta t \approx 11.2~$days into 512~MHz bins, and the C- and X-band observations at the time of the radio flare ($\delta t \approx 19.0~$days) into 256~MHz bins (chosen to maximize signal to noise while achieving the smallest width). The summary of this investigation can be found in Table~\ref{tab:VLA_epoch3_4_5_frequency}. We find no spectral evidence of DISS, with the frequency bins of all epochs and bands having flux densities consistent with either the measurement from the relevant sideband or across the entire observing band  (Figure~\ref{fig:DISS}, top, middle).

The effects of DISS can also manifest themselves as rapid variation in time. Thus, we search for temporal evidence of DISS at $\delta t \approx 19.0$~days, as this epoch has the highest signal to noise and is the most anomalous in terms of our afterglow modeling (Section~\ref{sec:fs_model}). To search for this temporal variation, we split the C- and X-band observations into 8 time bins of $\approx 510~$s each (the time between each phase calibrator cycle). The summary of this investigation can be found in Table~\ref{tab:VLA_epoch4_time}. We find no significant temporal variation in flux within either the C-band nor X-band observation (Figure~\ref{fig:DISS}, bottom), further indicating that DISS is not responsible for the radio flare at $\delta t \approx 19.0~$days.

The effects of scintillation are strongly frequency dependent, and far from the transition frequency, the variability timescales are longer and the maximum flux change is reduced. Therefore, we do not expect the detections obtained with MeerKAT to be affected by scintillation despite being in the strong scattering regime. The average observing length for our MeerKAT observations is 4\,hours, whereas RISS is expected to create flux modulation on a much longer timescale of 80\,hours at 1.3\,GHz, and DISS is expected to create flux modulation on a much shorter timescale of 0.2\,hours at 1.3\,GHz. Our long MeerKAT observations will average over the effects of DISS. Unfortunately, the signal to noise ratio of the epochs in which the afterglow is detected are too low to create shorter time chunks to search for evidence of scintillation. 

Therefore, we conclude that scintillation can not explain the radio flare of GRB\,210726A. We note that \citet{gbw+2018A&A...614A..29G} found scintillation was the only plausible explanation for the radio light curve of GRB\,151027B, despite flux changes of a factor of $> 5$. To explain the large flux modulation, they invoke a scattering screen at a smaller distance to the observer, multiple scattering screens, or a combination of the two. However, the thorough exploration of this scenario is beyond the scope of this work.

\subsection{Off-axis afterglow}

The rapid rise of the C-band afterglow of GRB\,210726A between  $\delta t \approx 11$ and $19~$days is reminiscent of an off-axis jet similar to GW\,170817 \citep[e.g.][see Figure~\ref{fig:Radio_Observations}]{margutti2018, alexander2018, fong2019}.
In the case of off-axis afterglows, the radiation emitted by the FS is not visible until the jet has decelerated to the point at which the observer is within the beaming cone of the jet. Off-axis afterglows are characterized by the comparatively late-time appearance of a rapidly rising broadband component which occurs at all wavelengths simultaneously \citep{granot2002, beniamini2022}.  However, the X-ray counterpart of GRB\,210726A does not show the same rapid rise as observed at C-band. Instead, it broadly consists of two broken power laws which are both decaying from early times, which is instead consistent with an on-axis jet. The lack of achromatic behaviour between the radio and X-ray wavebands rules out the possibility of the jet being viewed off-axis.

\subsection{Variable Circumburst Density}

The first scenario we consider arises as a result of possible variations in the circumburst environment which would be imprinted on the afterglow radiation \citep{wl2000ApJ...535..788W, dl2002ApJ...565L..87D}. This scenario was used to explain the achromatic optical variability of the long GRBs\,021004 and 050502A \citep{lrc+2002A&A...396L...5L, 2003NewA....8..495N, hp2003ApJ...586L..13H, 2005ApJ...630L.121G}, and also the optical/IR and radio light curve variability in the long GRB\,000301C \citep{bsf+2000ApJ...545...56B}. In this scenario, $F_{\nu} \propto n^{1/2}$ when $\nu_{\rm m}<\nu_{\rm obs}<\nu_{\rm c} $ and $F_{\nu, \rm max}$ follows the same dependency. Therefore, for GRB\,210726A, the apparent increase in $F_{\nu, \rm m}$ by a factor of $\approx 2$--$3$ would indicate an increase in $n_{0}$ by a factor of $\approx 4$--$9$.

However, this scenario fails in the context of GRB\,210726A for a number of reasons. First, we should also expect a flare at L-band, because self-absorbed radiation (i.e. $F_{\nu_{\rm sa}<\nu}$ and $\nu_{\rm sa}$) are both heavily dependent on $n_{0}$, yet the L-band light curve is relatively constant. Second, the example events listed above are all long GRBs, meaning they were produced by stellar explosions as opposed to NS-NS mergers. Circumburst density variations are easier to rectify within the stellar explosion scenario because the progenitors are massive stars with strong stellar winds and high mass loss rates. Changes in mass loss rate during the end phases of the star's life could easily change the density in the circumburst environment. However, it is difficult to create conditions for large changes in the circumburst density so close to the merger site in the NS-NS merger scenario. Finally, several studies have shown that density variations would result in little to no variability in the afterglow light curve, even when the density variation is significant \citep[e.g.][]{ng2007MNRAS.380.1744N, vemw+2009MNRAS.398L..63V, gvem2013ApJ...773....2G, gwl+2014ApJ...792...31G}, indicating that the sharp rise we observe is not physically motivated by this scenario. We therefore rule out a variable external density as the mechanism behind the radio flare. 

\newpage 
\bibliographystyle{apj}
\bibliography{library,journals_apj}

\clearpage

\end{document}